\documentclass[tighten, times, twocolumn]{aastex62}

\usepackage{ifthen}
\usepackage{amssymb}
\usepackage{amsmath}

\def\Teq{$T_{\rm eq}$}

\newcommand\Rpl{\ifmmode{R\sb{\rm pl}}\else$R\sb{\rm pl}$\fi}
\newcommand\Mpl{\ifmmode{M\sb{\rm pl}}\else$M\sb{\rm pl}$\fi}
\def\Re{\ensuremath{R_{\oplus}}}
\def\Me{\ensuremath{M_{\oplus}}}
\def\Mo{\ensuremath{M_{\odot}}}

\submitjournal{ApJ}

\shorttitle{Close-in sub-Neptunes reveal the past rotation history of their host stars}
\shortauthors{Kubyshkina et al.}

\begin{document}

\title{Close-in sub-Neptunes reveal the past rotation
  history of their host stars: atmospheric evolution of planets in the
  HD3167 and K2-32 planetary systems}

\correspondingauthor{Daria Kubyshkina}
\email{daria.kubyshkina@oeaw.ac.at}

\author[0000-0003-4426-9530]{D. Kubyshkina}
\affiliation{Space Research Institute, Austrian Academy of Sciences, Schmiedlstrasse 6, A-8042 Graz, Austria}

\author{P. E. Cubillos}
\affiliation{Space Research Institute, Austrian Academy of Sciences, Schmiedlstrasse 6, A-8042 Graz, Austria}

\author{L. Fossati}
\affiliation{Space Research Institute, Austrian Academy of Sciences, Schmiedlstrasse 6, A-8042 Graz, Austria}

\author{N. V. Erkaev}
\affiliation{Institute of Computational Modeling of the Siberian Branch of the Russian Academy of Sciences, 660036 Krasnoyarsk, Russian Federation}
\affiliation{Siberian Federal University, 660041 Krasnoyarsk, Russian Federation.}

\author{C. P. Johnstone}
\affiliation{Institute for Astronomy, University of Vienna, T\"urkenschanzstrasse 17, A-1180 Vienna, Austria}

\author{K. G. Kislyakova}
\affiliation{Institute for Astronomy, University of Vienna, T\"urkenschanzstrasse 17, A-1180 Vienna, Austria}
\affiliation{Space Research Institute, Austrian Academy of Sciences, Schmiedlstrasse 6, A-8042 Graz, Austria}

\author{H. Lammer}
\affiliation{Space Research Institute, Austrian Academy of Sciences, Schmiedlstrasse 6, A-8042 Graz, Austria}

\author{M. Lendl}
\affiliation{Space Research Institute, Austrian Academy of Sciences, Schmiedlstrasse 6, A-8042 Graz, Austria}

\author{P. Odert}
\affiliation{IGAM/Institute of Physics, University of Graz, Universit\"atsplatz 5, A-8010 Graz, Austria}
\affiliation{Space Research Institute, Austrian Academy of Sciences, Schmiedlstrasse 6, A-8042 Graz, Austria}

\author{M. G\"{u}del}
\affiliation{Institute for Astronomy, University of Vienna,
T\"urkenschanzstrasse 17, A-1180 Vienna, Austria}
%
\begin{abstract}
\noindent Planet atmospheric escape induced by high-energy stellar
irradiation is a key phenomenon shaping the structure and
evolution of planetary atmospheres. Therefore, the present-day
properties of a planetary atmosphere are intimately connected with
the amount of stellar flux received by a planet during its
lifetime, thus with the evolutionary path of its host star. Using
a recently developed analytic approximation based on hydrodynamic
simulations for atmospheric escape rates, we track within a
Bayesian framework the evolution of a planet as a function of
stellar flux evolution history, {constrained by the measured
planetary radius, with the other system parameters as priors.} We
find that the ideal objects for this type of study are close-in
sub-Neptune-like planets, as they are highly affected by
atmospheric escape, and yet retain a significant fraction of their
primordial hydrogen-dominated atmospheres. Furthermore, we apply
this analysis to the HD3167 and K2-32 planetary systems. For
HD3167, we find that the most probable irradiation level at 150
Myr was between 40 and 130 times solar, corresponding to a
rotation period of $1.78^{+2.69}_{-1.23}$ days. For K2-32, we find
a surprisingly low irradiation level ranging between half and four
times solar at 150 Myr. Finally, we show that for multi-planet
systems, our framework enables one to constrain poorly known
properties of individual planets.
\end{abstract}

%
\section{Introduction}\label{sec:introduction}
{For late-type stars (i.e., later than mid F-type)}, rotation rate
and high-energy radiation (X-ray and EUV below 912\,\AA; hereafter
XUV) are intimately connected, with faster rotating stars being
XUV brighter
\citep[e.g.,][]{pallavicini1981,pizzolato2003,johnstone2015rot}.
During their main-sequence lifetime, the rotation rate and XUV
flux of late-type stars decrease with time. This evolution does
not follow a unique path, as stars born with the same mass and
metallicity can have widely different initial rotation rates, thus
XUV fluxes \citep[e.g.,][]{mamajek2008,johnstone2015rot}. This
non-uniqueness holds up to {about 2\,Gyr}, at which point the
evolutionary tracks of both rotation rate and XUV emission
converge to one path \citep[e.g.,][]{tu2015}. This implies that,
for stars older than {about 2\,Gyr}, it is not possible to infer
their past XUV emission from their present stellar properties.

This problem becomes relevant when trying to understand the
evolution of planetary atmospheres. The XUV stellar flux is
absorbed in the upper atmosphere of planets leading to atmospheric
expansion and eventually escape, which has been shown to occur
efficiently on close-in Jupiter- and Neptune-mass planets
\citep[e.g.,][]{vidal2003,fossati2010,lecavelier2012,ehrenreich2015}.
Furthermore, atmospheric escape plays a significant role in
shaping the observed exoplanet population and atmospheric
properties \citep[e.g.,][]{lopez2013, lopez2014, jin2014,
tian2015, owen2017, owen2018, jin2018}. Since in most cases
atmospheric mass-loss rates are directly proportional to their XUV
irradiation, escape is most effective when planets are young, thus
the importance of knowing the past evolution of the stellar XUV
radiation.

Luckily, for stars hosting close-in planets with
hydrogen-dominated atmospheres, we can infer the stellar XUV
irradiation tracks by modeling the planets' atmospheric evolution.
The amount of XUV flux received by a planet during its lifetime
determines the amount of hydrogen remaining in its envelope: since
the radius of planets with masses below 0.35 Jupiter masses
strongly correlates with its hydrogen envelope content
\citep[e.g.,][]{lopez2014,hatzes2015}, the currently observed
planetary radius constrains the past evolution of the stellar XUV
emission.

Here, we present a framework, based on mass-loss rates extracted
from hydrodynamic simulations \citep{kubyshkina2018grid},
{enabling the modeling of} the atmospheric evolution of
super-Earths and sub-Neptunes that we apply to the HD3167 and
K2-32 planetary systems. Section~\ref{sec:method} summarises the
algorithms and tools employed to compute the planetary atmospheric
evolutionary tracks. Section~\ref{sec:initial_conditions} shows
the results of tests conducted on mock systems.
Section~\ref{sec:results} presents the results {obtained for the
HD3167 and K2-32 planetary systems}, while
Section~\ref{sec:conclusion} gives our discussion and conclusions.

\section{Atmospheric Evolution Model}
\label{sec:method}
To model the evolution of planetary atmospheres, we employ the
scheme described by \citet{kubyshkina2018grid}. Starting from a
set of initial system parameters, as a planet loses mass from its
hydrogen envelope, at each time-step the code extracts the
atmospheric mass-loss rate from an analytical approximation based
on a grid of {one dimensional} hydrodynamic models, uses this
value to update the atmospheric mass fractions, and estimates the
planetary radius, which we then compare to the observed radius
once the age of the system has been reached. Here, we summarise
the key steps and describe the applied improvements.

There are three key ingredients: a model of the stellar flux
evolutionary track, a model relating planetary parameters and
atmospheric mass, and a model computing atmospheric escape rates.

\begin{figure}[t]
\includegraphics[width=\hsize]{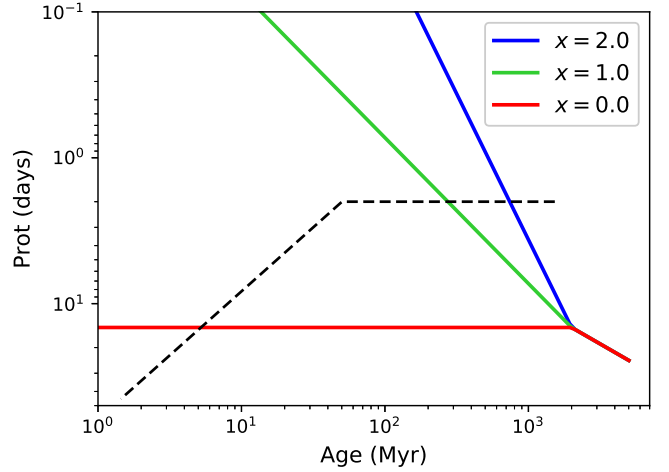}
\caption{{Model rotation-period curves as a function of age for
    a range of $x$ values (colored solid curves).  The black dashed
    curve denotes the time-dependent threshold when a star drops out
    of saturation, \citep[see Equation~(2) and Figure~2 of ][]{tu2015}.}}
\label{fig:prot}
\end{figure}

For late-type stars, the stellar XUV flux out of the saturation
regime depends on the stellar mass and rotation period, where the
latter is time-dependent. \citet{mamajek2008} suggested the
following analytical approximation for the evolution of the
stellar rotation period (in days)
\begin{equation}
\label{eq:Trot}
P_{\rm rot}(\tau) = 0.407\,[(B-V)_0-0.495]^{0.325}\,\tau^{0.566},
\end{equation}
where $\tau$ is the stellar age (in Myr), and $(B-V)_0$ is the
reddening-free stellar color. Equation~(\ref{eq:Trot}) represents
the average approximation based on a large set of late-type
dwarfs.  However, the evolutionary tracks for stellar rotation
rates are non-unique \citep{johnstone2015rot,tu2015}. To account
for the different rotation-rate histories of different stars
(i.e., from slow to fast rotator), we model the rotation period as
a power law in $\tau$, normalised such that the rotation period at
the present age ($T_{\rm age}$) is consistent with the measured
stellar rotation period ($P_{\rm rot}^{\rm now}$), obtaining

\begin{equation}
\label{eq:Trot1} P_{\rm rot} =
    \begin{cases}
    P_{\rm rot}^{\rm now} \left(\frac{\tau}{T_{\rm age}}\right)^{0.566}\,, & \text{$\tau\geq 2$~Gyr}\\
    P_{\rm rot}^{\rm now} \left(\frac{2 {\rm Gyr}}{T_{\rm age} {\rm [Gyr]}}\right)^{0.566}\, \left(\frac{\tau {\rm [Gyr]}}{2 {\rm Gyr}}\right)^{x}\,,& \text{$\tau<2$~Gyr}
    \end{cases}
\end{equation}
{where the exponent $x$ is a positive value, typically ranging
between 0 and $\sim$2, controlling the stellar rotation period at
ages younger than 2\,Gyr. Figure \ref{fig:prot} shows examples of
the parametric $P_{\rm rot}$ curves for a range of $x$ values
equal to 0, 1, and 2.}

We then derive the stellar X-ray and EUV luminosity from the
rotation period following \citet{pizzolato2003} and
\citet{wright2011} that relate rotation rates and stellar masses
to X-ray luminosity, including saturation effects for the fastest
rotators, and \citet{sanz2011} relating X-ray and EUV fluxes. To
account for variations of the stellar bolometric luminosity with
time, we use the MESA Isochrones and Stellar Tracks
\citep[MIST,][]{paxton2018} model grid.

To estimate the planetary mass-loss rate at a given time, we use
the analytic formulas provided by \citet{kubyshkina2018approx},
which are based on a grid of one-dimensional hydrodynamic
upper-atmosphere models \citep{kubyshkina2018grid}. The analytic
approximation requires as input stellar XUV flux, planetary mass
(\Mpl), radius (\Rpl), equilibrium temperature (\Teq), and orbital
separation ($d_0$).

To estimate the atmospheric mass of a planet $M_{\rm atm}$ as a
function of planetary mass, radius, and \Teq, we pre-compute
$M_{\rm atm}$ following \citet{johnstone2015} for the range of
parameters used in \citet{kubyshkina2018approx}, among which we
further interpolate during an evolution run.  For a given core
mass, we determine the core radius $R_{\rm core}$ assuming an
Earth-like density, as for planets with hydrogen-dominated
envelopes the core composition has little influence on planetary
size \citep[e.g.,][]{lopez2014,petigura2016}.

We combine these ingredients to compute the planetary atmospheric
evolutionary tracks.  For each run, we assume that the planetary
orbital separation and stellar mass remain constant, and ignore
the contribution of gravitational contraction and radioactive
decay on the planetary \Teq\ during the first phases of evolution.
We further assume that every planet accreted a hydrogen-dominated
atmosphere from the proto-planetary nebula. We begin our
simulations at 5\,Myr, which is the typical lifetime of
protoplanetary disks \citep{mamajek2009}. We also consider an
initial planetary radius corresponding to a value of the
restricted Jeans escape parameter $\Lambda$ (computed for the
planetary core mass and \Teq\ at the beginning of the simulation)
of 5, where \citep{jeans1925,fossati2017}
\begin{equation}
\label{eq:lambda}
\Lambda = \frac{G M_{\rm p}m_{\rm H}}{k_{\rm b}T_{\rm eq}R_{\rm p}}\,,
\end{equation}
with $G$ gravitational constant, $m_{\rm H}$ mass of an hydrogen
atom, and $k_{\rm b}$ Boltzmann constant. We come back to this
point in Section~\ref{sec:initial_conditions}.

At each step of the evolution, first we compute the mass-loss rate
based on the stellar flux and system parameters, which we use to
update the atmospheric mass fraction, and the planetary radius. We
adjust the time step, such that the change in atmospheric mass is
less than 5\% of $M_{\rm atm}$.

Finally, we use a Bayesian approach to constrain the evolutionary
track of the stellar XUV luminosity fitting the currently observed
planetary radius. To this end, we combine the planetary evolution
model with the open-source Markov-chain Monte Carlo (MCMC)
algorithm of \citet{cubillos2017}, to compute the posterior
distribution for the stellar rotation rate and the considered
system parameters.

For each MCMC run, we let the planetary mass, age of the system,
present-time rotation period, orbital separation, and stellar mass
as free parameters, with Gaussian-like priors according to the
measured values and uncertainties. Within our scheme, we adopt a
stellar radius from the MESA evolutionary tracks, which depends on
the stellar mass. Similarly, we compute {\Teq} derived from the
adopted stellar radius and $d_0$, assuming zero Bond albedo and
full energy redistribution. The stellar radius may also be
implicitly included in the computation in case only an upper limit
for $P_{\rm rot}^{\rm now}$ has been given. {We parameterise the
stellar rotation evolution using $x$ of Equation\,\ref{eq:Trot1},
adopting a uniform prior ranging between 0 and 2.  However, we
provide the results in terms of the stellar rotation period at the
age of 150\,Myr, because this is more intuitive than $x$ and it
allows one to compare the results with the distribution of stellar
rotation periods measured in young open clusters.}

\begin{figure}[t]
\includegraphics[width=\hsize]{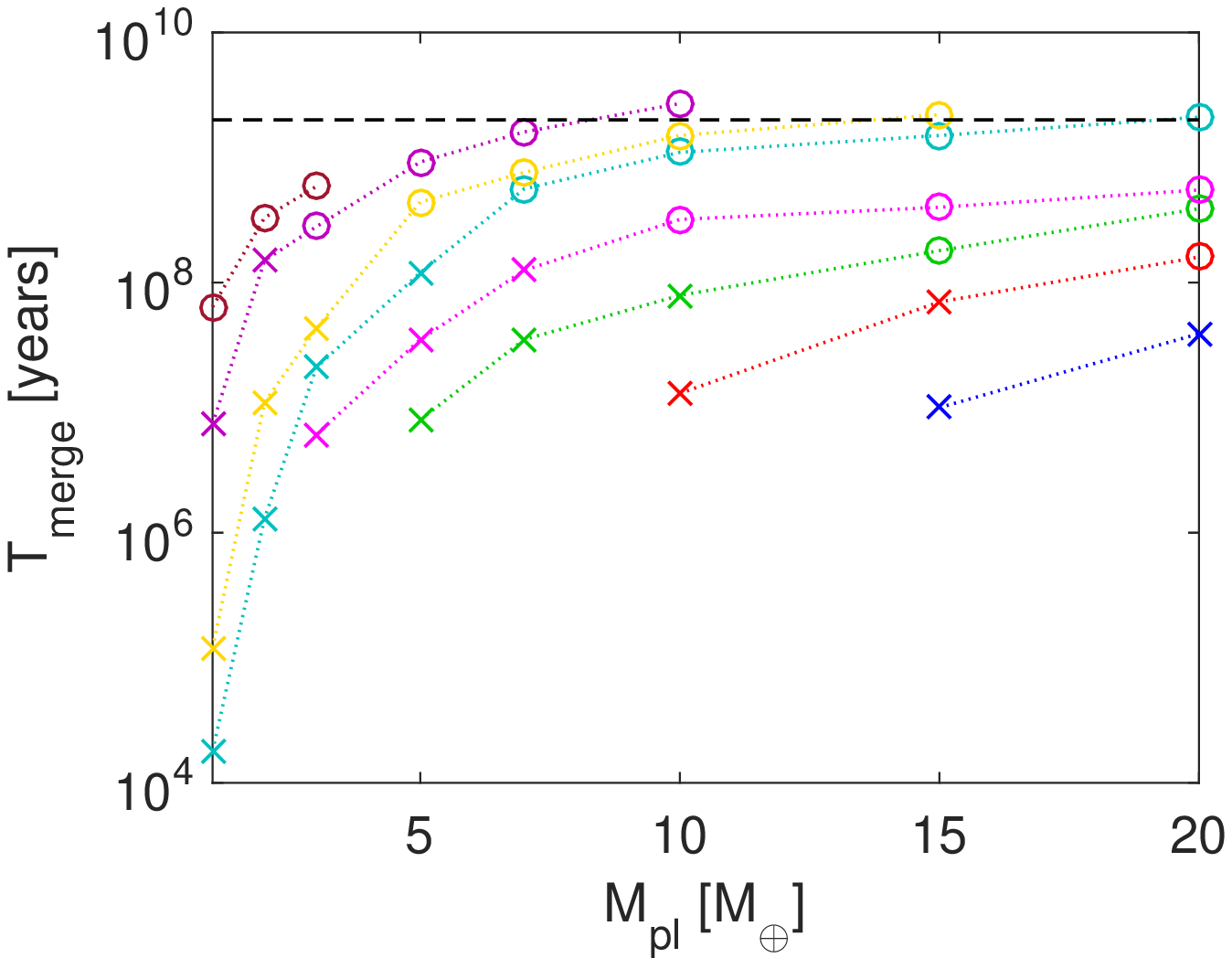}\\
\includegraphics[width=\hsize]{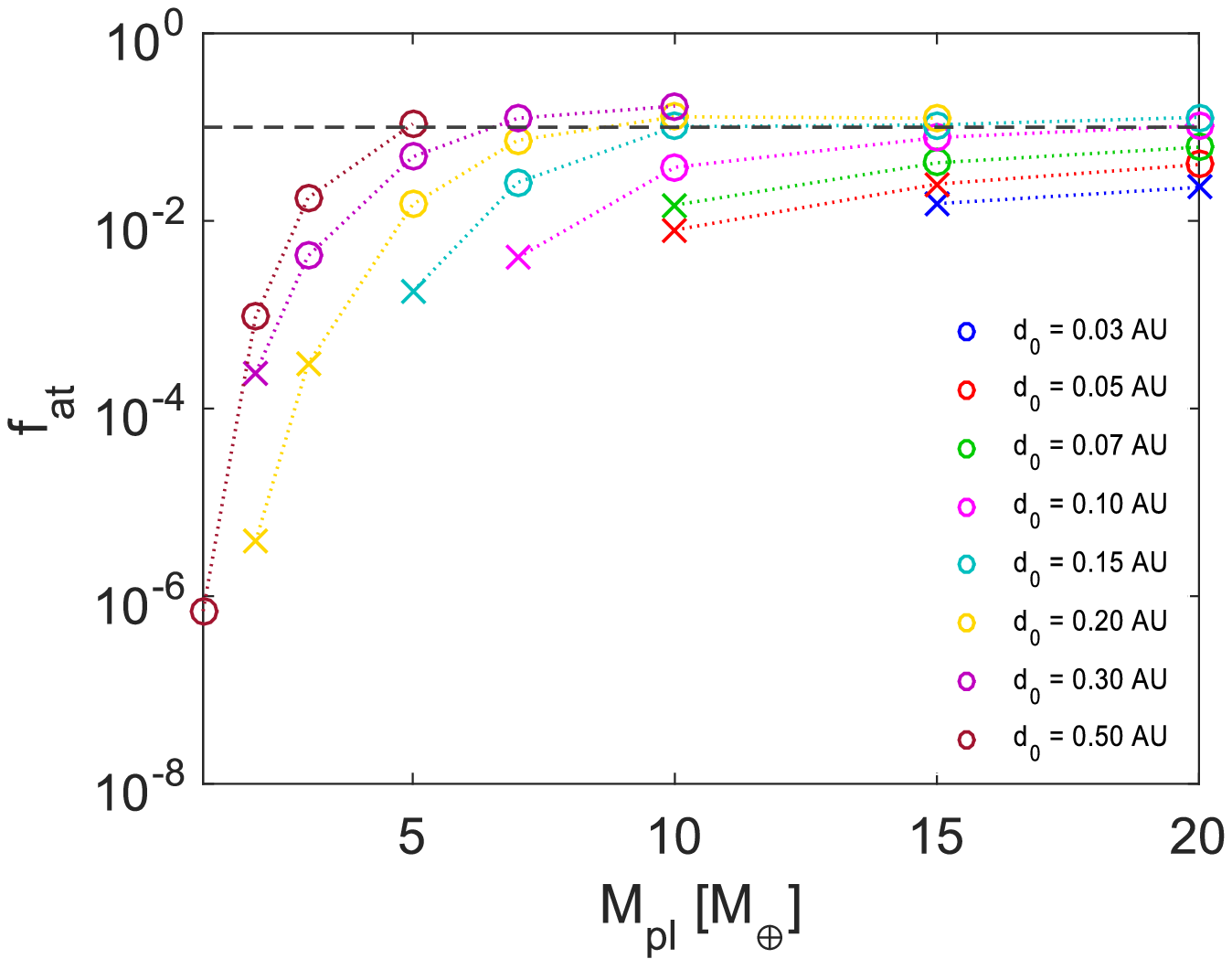}
\caption{Top: time at which the planetary atmospheric evolutionary
tracks merge as a function of \Mpl.  Bottom: minimum initial
atmospheric mass fraction for which the planetary atmospheric
evolutionary tracks merge as a function of \Mpl. Colors indicate
different orbital separations (see legend).  Crosses and circles
indicate whether the modelled planet loses or preserves a
hydrogen-dominated atmosphere within 10\,Gyr, respectively.  For
reference, the black dashed lines indicate in the top panel an age
of 2\,Gyr and in the bottom panel an atmospheric mass fraction of
10\%.} \label{fig:test}
\end{figure}

\begin{table*}
\begin{center}
\caption{{Radii and atmospheric mass fractions following a
forward-model evolution for the test planets.}}\label{tab:test}
\begin{tabular}{c|c|c|c}
  \hline
   & \multicolumn{3}{|c}{$R_{\rm pl}$ [\Re] / $f_{\rm at}$ [$\%$]} \\
  \cline{2-4}
  planet & slow & medium & fast \\
  \hline
  b & $3.44\pm 0.17$ / $2.3\pm0.8$ & $2.80\pm0.14$ / $0.2_{-0.2}^{+0.8}$ & $2.71\pm 0.14$ / $0.0_{-0.0}^{+0.9}$ \\
  c & $5.26\pm 0.26$ / $11.7_{-1.0}^{+1.2}$ & $4.12\pm 0.21$ / $6.9_{-1.3}^{+1.2}$ & $3.22\pm 0.16$ / $3.1_{-1.3}^{+1.1}$ \\
  d & $6.53\pm 0.33$ / $20.4_{-1.0}^{+0.3}$ & $4.53\pm 0.23$ / $11.0_{-0.8}^{+0.4}$ & $3.40\pm 0.17$ / $5.7_{-0.7}^{+0.5}$ \\
  \hline
\end{tabular}
\end{center}
\end{table*}

\section{Validation}
\label{sec:initial_conditions}
\subsection{Initial conditions}
\citet{kubyshkina2018grid} showed that for planets less massive
than 6\,\Me\ and orbital separations smaller than 0.1\,AU, the
initial radius has no impact on the final radius at the end of the
evolution. This is because, for a given \Mpl\ and $d_0$, a larger
(smaller) radius, hence larger (smaller) $M_{\rm atm}$, leads to a
higher (lower) mass-loss rate. Here, we look more thoroughly for
the \Mpl\ and $d_0$ values above which the atmospheric evolution
drawn by our framework starts to become sensitive to the initial
radius.

The main parameters controlling atmospheric escape at a given XUV
irradiation and \Rpl\ are \Mpl\ and $d_0$.  Therefore, we consider
systems composed of a 0.8\,\Mo\ medium rotator ($x = 1$) and
planets with masses ranging between 1 and 20\,\Me\ and orbital
separations ranging between 0.02 and 0.50\,AU.  For each
\Mpl--$d_0$ pair, we computed a series of evolutionary tracks
over a range of initial planetary radii, looking for the time at
which the evolutionary tracks merge ($T_{\rm merge}$), namely the
atmospheric mass fractions become the same within 2\%, and the
minimum initial radius for which this happens.

We set the upper limit of the tested initial radii at 10\,\Re, the
maximum \Rpl\ available in the grid of upper atmosphere models
\citep{kubyshkina2018grid}. We set the lower limit of the tested
initial radii based on the minimum expected atmospheric mass
fraction $f_{\rm at} = M_{\rm atm}/M_{\rm p}$. Since within the
protoplanetary disc the mass of the initially accreted envelope
grows with increasing planetary mass, we expect $f_{\rm at} >
10^{-4}$ for \Mpl\,$\leq$\,5\,\Me\ and $f_{\rm at} = 10^{-2}$ for
higher masses \citep{steokl2015}.

Figure~\ref{fig:test} shows that $T_{\rm merge}$ increases with
\Mpl\ and $d_0$, and that it remains always below $\sim$2\,Gyr.
Figure~\ref{fig:test} further shows the minimum atmospheric mass
fraction $f_{\rm at,min}$ required for the evolutionary tracks to
merge as a function of \Mpl\ and $d_0$. Both $T_{\rm merge}$ and
$f_{\rm at,min}$ increase with increasing \Mpl\ and $d_0$. For
systems with \Mpl\ above 5\,\Me\ and $d_0$ larger than 0.2\,AU,
$f_{\rm at,min}$ lies above the atmospheric mass fraction that can
be expected to be originally accreted by each planet before the
dispersal of the protoplanetary disc based on \citet{steokl2015}.
Therefore, the results of the atmospheric evolution drawn by our
framework start to become sensitive to the initial radius for
planets more massive than 5\,\Me\ and lying at orbital separations
larger than 0.2\,AU.

When applied to real cases, the results presented in
Figure~\ref{fig:test} should be considered as indicative. This is
because the mass and rotation rate, thus XUV emission, of a given
host star are possibly different from the values considered here.
However, we can conclude that for sub-Neptune-like planets within
0.2\,AU the initial conditions are in general not important, while
for heavier and/or more distant planets this is not the case.  We,
however, evaluate in each case the impact of the choice of the
initial conditions on the results.

\subsection{Reproducibility of mock systems}
\label{sec:retrievaltest}

To test whether and in which cases the retrieval approach can
constrain the stellar XUV luminosity evolutionary tracks, we carry
out an injection-retrieval test. To do so, we create a 5\,Gyr old
system, with a stellar mass ($M_*$) of 0.8\,\Mo\ and $P_{\rm
rot}^{\rm now} = 30$\,days.  We simulate three planets (labeled ``b'',
``c'', and ``d'') with masses of 20, 15, and 8\,\Me\ at orbital
separations of 0.05, 0.1, and 0.2\,AU, respectively. These
parameters make sure that the initial planetary radius, set as the
value for which $\Lambda = 5$, has no influence on the results.
The core radii of the three planets are 2.71, 2.47, and 2.00\,\Re,
respectively.

\subsubsection{Typical uncertainties}

To estimate the system-parameter uncertainties, we look at the
typical measured uncertainties listed in the NASA Exoplanet
Archive\footnote{{\tt http://exoplanetarchive.ipac.caltech.edu}}.
For planets similar to those considered in this test (i.e., \Mpl\
below 25\,\Me, excluding upper limits, \Teq\ between 300 and
2000\,K, stellar mass between 0.4 and 1.3\,\Mo), we adopt stellar
mass uncertainties of 4\% (signal-to-noise ratio, S/N, of 25),
orbital separation uncertainties of 1.5\% (S/N\,=\,66.7),
planetary mass uncertainties of 20\% (S/N\,=\,5), and stellar age
uncertainties of 30\% (S/N\,=\,3.3). Note that the estimated age
uncertainties vary significantly; however, this uncertainty has
the largest influence only for systems much younger than the focus
of this work (see below).

Finally, for the measured rotation period it is not possible to
reliably associate a typical uncertainty because the number of
stars for which $P_{\rm rot}^{\rm now}$ has been measured is
small, and in most cases $P_{\rm rot}^{\rm now}$ either
corresponds just to an upper limit or has been estimated from
approximations, such as that of \citet{mamajek2008}. On the basis
of the few stars for which $P_{\rm rot}^{\rm now}$ has been
measured, we adopt an uncertainty of 15\% (S/N\,$\approx$\,6.7).

\subsubsection{Results}

We simulate three evolution scenarios with $x$\,=\,0.1, 0.566,
and 1.5, corresponding to an initially ``slow'', ``medium'', and
``fast'' rotator or alternatively rotational periods at 150\,Myr
($P_{\rm rot}^{150}$) of 13.8, 4.1, and 0.4 days, respectively.
The planetary radii and atmospheric mass fractions (accounting for
the uncertainties on the parameters) obtained for these scenarios
and for planets ``b'', ``c'', and ``d'' following a forward model
are listed in Table~\ref{tab:test}. When orbiting a medium or a
fast rotator, planet ``b'' loses (almost) completely its
atmosphere as, within the uncertainties, its atmospheric mass
fraction remains below 1.0\%.

{For all measured/derived system parameters we set a gaussian-like
prior defined by the injected values and their uncertainties. For
the parameter $x$, which defines the evolution of the stellar
rotation period, we set a uniform prior between 0 and 2.}

\begin{figure*}
\begin{center}
  \includegraphics[width=0.8\hsize]{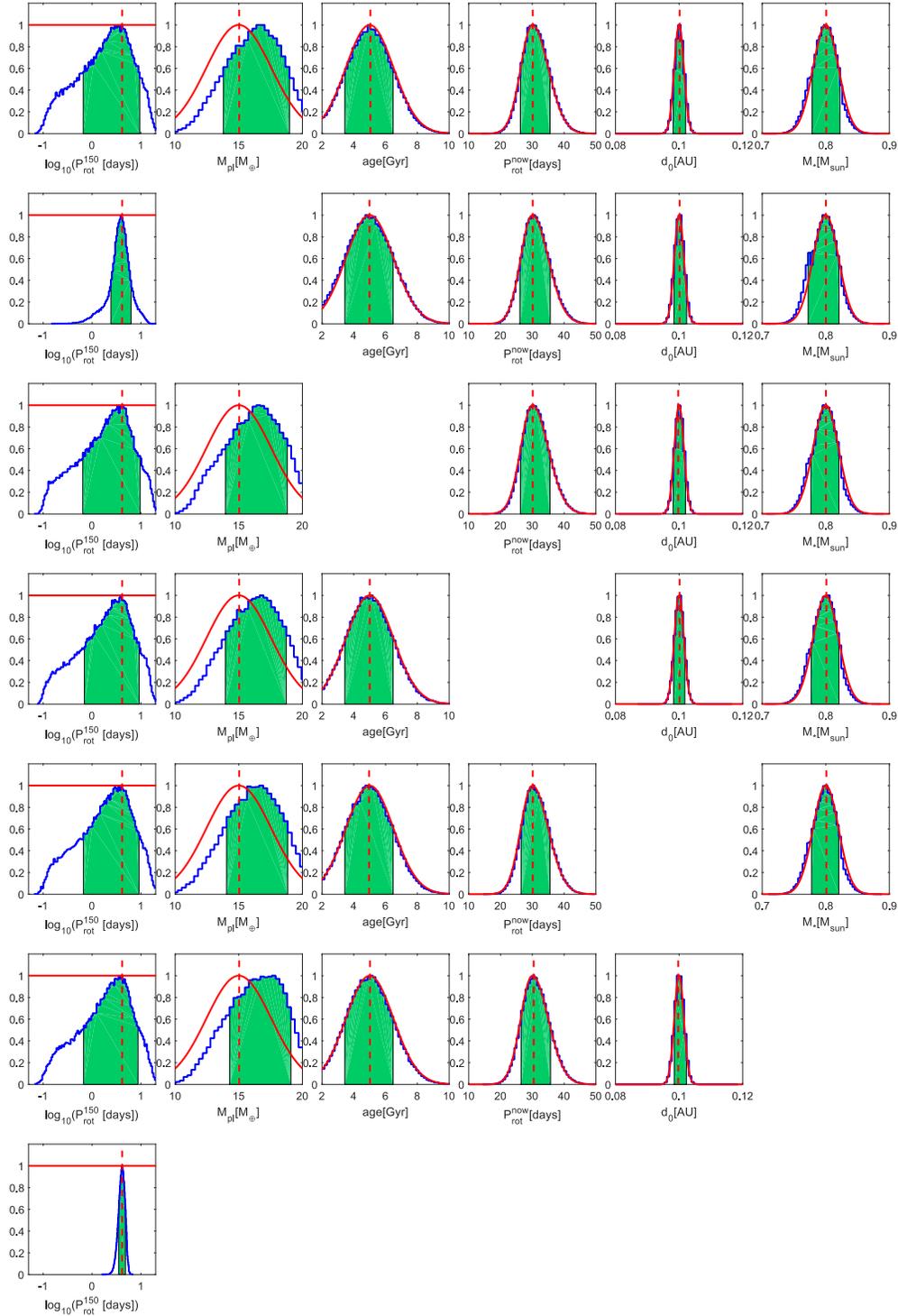}\\
  \caption{Posterior distributions for the injection-retrieval runs described in Section~\ref{sec:retrievaltest} for the test planet ``c'' orbiting a
  medium rotator.  From left to right, each column shows the posterior distribution for the stellar rotation period at 150\,Myr ($P_{\rm rot}^{150}$),
  planetary mass, age of the system, present-time rotation period, orbital separation, and stellar mass, respectively.  From top to bottom, each row shows
  the results assuming all system parameters free, and then fixing one of these parameters at a time, respectively.  The missing panels indicate the parameter
  that has been fixed in each run.  The bottom row, composed by a sole panel, shows the posterior stellar rotation period at 150\,Myr while fixing all input
  parameters, {except for the stellar rotation period}.  The green shading denotes the 68\% highest posterior density (HPD) credible intervals.  {The red dashed vertical
  lines show the injected values,} whereas the red solid lines show the prior distributions of the input parameters.}
\label{fig:test_hist}
\end{center}
\end{figure*}

As a typical example of the results obtained for the test
systems, Figure~\ref{fig:test_hist} shows the MCMC posterior
distributions for the case of the ``c'' planet orbiting the medium
rotator. The results obtained for the other tests are shown in the
Appendix (Figures~\ref{fig:appx1}--\ref{fig:appx8}).

In Figure~\ref{fig:test_hist}, the top row shows the posterior
distribution obtained when all system parameter were set free. To
identify whether there are dominant parameters driving the results
and which ones they are, we repeated the analysis fixing one
parameter at a time.  For completeness, we run a test fixing all
parameters except for {the stellar rotation period at 150\,Myr},
recovering the injected $P_{\rm rot}$ distribution.

For each modelled case, the posterior distribution for the
stellar rotation period matches the injected value within the 68\%
highest posterior density (HPD) credible interval. Our tests
indicate that planetary mass is the main parameter influencing the
posterior distribution of the rotation period, with the other
parameters playing a minor role.

\begin{figure*}
\begin{center}
  \includegraphics[width=\hsize]{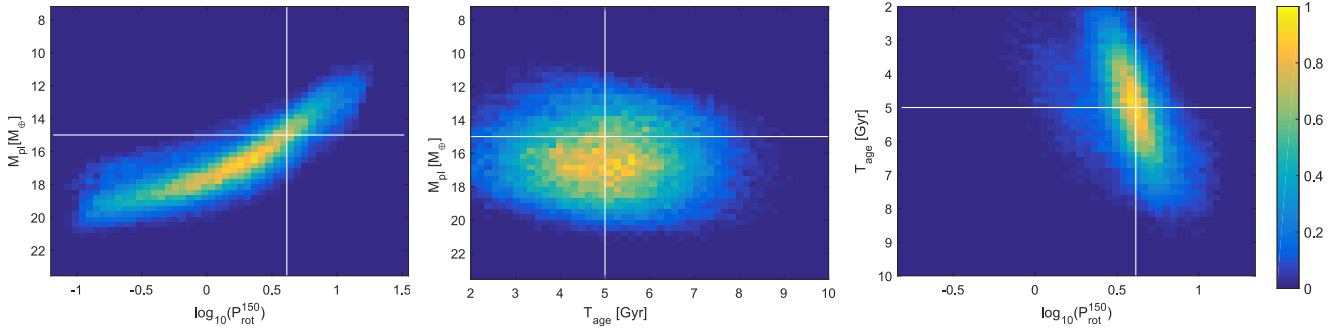}\\
  \caption{Pair-wise posterior distributions of $P_{\rm rot}^{150}$ vs \Mpl\ (left), system's age vs  \Mpl\ (middle), and $P_{\rm rot}^{150}$ vs system's age, the latter with \Mpl\ fixed (right), for the test planet ``c'' orbiting the medium rotator. The white solid lines indicate the injected values.}\label{fig:test_images}
\end{center}
\end{figure*}

Our tests also show that the posterior distribution of all
parameters, except for the planetary mass, are consistent with the
input distribution.  For the planetary mass, we find that the
posterior distributions are consistently shifted towards higher
values than their prior distributions; however, the results are
consistent within 1$\sigma$ with the prior. The pair-wise
posterior distributions (Figure~\ref{fig:test_images}) reveal a
well-defined degeneracy between $P_{\rm rot}^{150}(M_{\rm pl})$
and $M_{\rm pl}$, which occurs because one can arrive at the same
planetary radius through a lower (higher) planetary mass and a
slower (faster) rotator, thus lower (higher) past XUV irradiation.
This explains the shift between prior and posterior distributions.
This effect is more evident for closer-in planets and for slower
rotators. In the case of the slow-rotator simulation, we also find
a slight shift of the posterior of the stellar rotation period
compared to the injected value, but still lying well within
1$\sigma$.  This indicates that the algorithm prefers to reach the
final planetary radius with a slightly heavier planet and a
slightly faster rotator.

For the planets orbiting the fast rotator, the posteriors for
planetary mass are narrower than the prior.  This happens also for
the test planet ``b'', which is the closer-in of the three,
orbiting the medium rotator (Figure \ref{fig:appx4}), for which
the final atmospheric mass fraction is below 1\%. This happens
because the thin atmosphere remaining at the end of the simulation
can be obtained only through a narrow range of planetary masses.
This indicates that our model imposes constraints on the plausible
physical parameters of the system, leading to constraints on the
planetary masses.

The left and middle panels of Figure~\ref{fig:test_images} show
the pair-wise posterior distributions of $P_{\rm rot}^{150}$ vs.
$M_{\rm pl}$ and $T_{\rm age}$ vs. $M_{\rm pl}$, for the
medium-rotator case when all parameters are free. The correlation
is rather narrow, meaning that, despite the whole range of
possible stellar rotation rates is wide, at each given planetary
mass the range of possible stellar rotation rates is narrow. For
the case of the slow rotator, we also find a correlation between
\Mpl\ and $T_{\rm age}$.  This occurs because the atmospheric mass
fraction changes more slowly over time for the slow rotator. A
correlation between $P_{\rm rot}^{150}$ and $T_{\rm age}$ is the
direct consequence of the other two correlations (right panel of
Figure~\ref{fig:test_images}). The other system parameters do not
show significant correlations.

For some of the test planets, such as planets ``b'' and ``c''
orbiting the fast rotator, we obtained broad posterior
distributions, thus weak constraints on the stellar rotation rate
history. The case of planet ``b'', the closer-in and most massive
planet, orbiting the fast or medium rotator lead to the (almost)
complete escape of the atmosphere. Therefore, even though formally
the injected value of the rotation period lies within the 68\% HPD
credible interval, we can just obtain an upper limit for the
stellar rotation period. Our tests  indicate that for planets with
a thin hydrogen envelope, a meaningful use of our scheme requires
a high accuracy of the system parameters: the thinner the hydrogen
atmosphere, the higher the required accuracy on the system
parameters.

To make sure that our results do not depend on the adopted masses
and positions of the test planets, we carry out additional runs
considering test planets orbiting at distances from the host star
ranging between 0.1 and 0.25\,AU and having masses ranging between
5 and 20\,\Me. The results of these tests, not shown here, are
consistent with those presented above, confirming that the
accuracy with which we are able to constrain the stellar rotation
period depends mainly on the accuracy of the others parameters
(predefined outside of our framework) and the uncertainty of the
data.

%

%
\subsubsection{Influence of the uncertainties}
%
\begin{figure}[t]
  \includegraphics[width=\hsize]{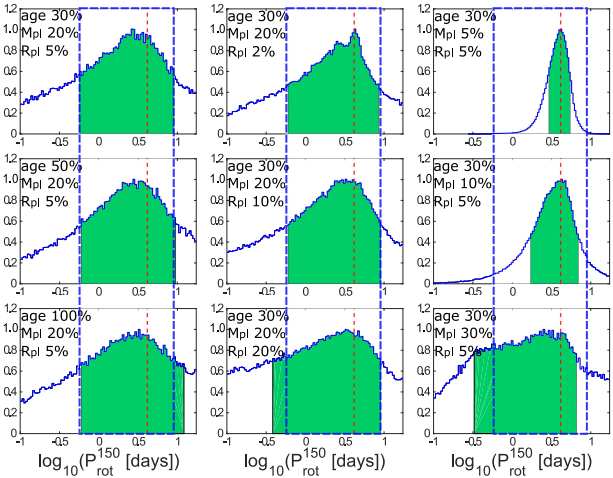}\\
  \caption{Posterior distribution of the rotation period at an age of 150\,Myr for the test planet ``c'' orbiting the medium rotator. The different panels
  present the results obtained following different assumptions on the uncertainties of {the system age, planetary mass, and planetary radius}, as labelled on the top-left corner
  of each panel. The blue dashed rectangles show for reference the width of the 68\% HPD credible interval for the test reference uncertainties (top-left panel).}\label{fig:test_unc}
\end{figure}

From the test runs, illustrated in Figures~\ref{fig:test_hist} and
\ref{fig:appx1}--\ref{fig:appx8}, we expect that the main
uncertainties affecting our ability to constrain the stellar
rotation history are those on \Mpl\ and \Rpl. Therefore, we
performed additional tests to examine how the results change by
changing the uncertainties on these {values}. For the planetary
radius, we varied the uncertainty between 2 and 50\%, while for
the planetary mass, we varied the uncertainty between 5 and 50\%.
In addition, because of the large spread with which ages of
planetary systems are derived, we present the results of
additional runs considering age uncertainties of 50 and 100\%.

Figure~\ref{fig:test_unc} shows the results of these tests. The
top-left panel is the reference distribution, which coincides with
the top-left panel of Figure~\ref{fig:test_hist}. The first,
second, and third columns present the results obtained varying the
uncertainties on the system's age, planetary radius, and planetary
mass, respectively. There is little variation in the resulting
posterior distributions for $P_{\rm rot}^{150}$ when varying the
system's age uncertainty, because the main changes in atmospheric
mass fraction occur during the early stages of evolution---before
2\,Gyr.  Therefore, the uncertainty on the system's age plays a
role only for very young systems.  As expected, increasing the
uncertainty on the planetary radius and mass broadens the
posterior distribution of $P_{\rm rot}^{150}$, becoming nearly
flat for a radius and/or mass uncertainties of 50\% (not shown).
Improving the planetary-mass uncertainty has the largest impact on
the expected $P_{\rm rot}^{150}$ posterior distribution.  For the
real planets examined in this study, the uncertainties on
planetary mass are 13--16\%.

\subsubsection{Simultaneous modelling of multiple planets in the same
system}\label{sec:test:multiplanet}
%
\begin{figure*}
\begin{center}
  \includegraphics[width=\hsize]{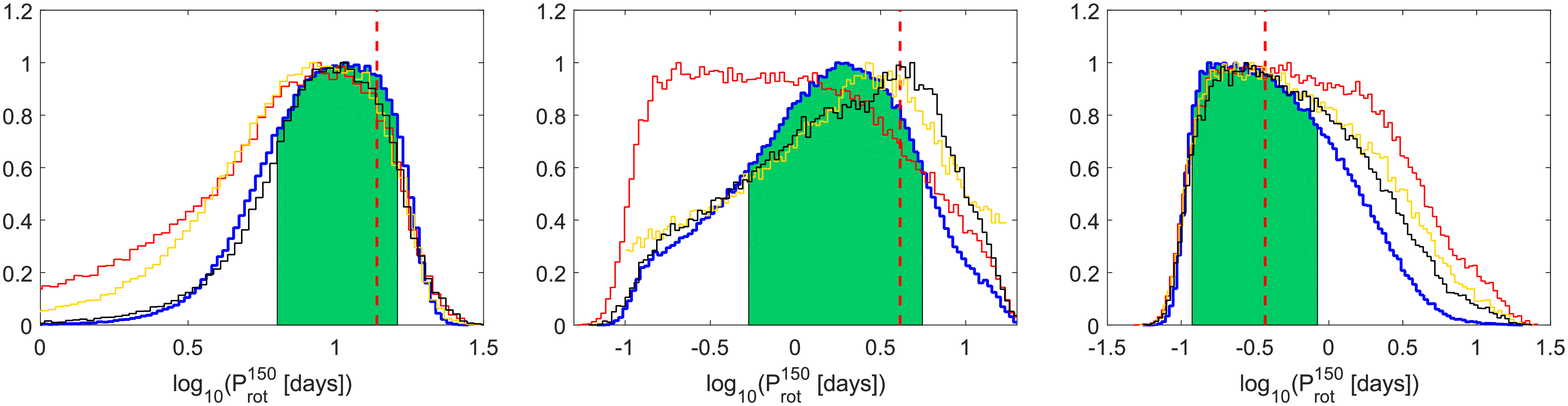}\\
  \caption{Results of the simultaneous modelling of the three test planets orbiting a slow (left), medium (middle), and fast (right) rotator. The red, yellow, and black thin lines show the posterior distributions obtained by modelling separately the ``b'', ``c'', and ``d'' planets, respectively.}\label{fig:test_sys}
\end{center}
\end{figure*}

Here we explore what can be obtained by analysing simultaneously
two or more planets orbiting the same star.
Figure~\ref{fig:test_sys} shows the posterior distributions for
$P_{\rm rot}^{150}$ obtained by simultaneously modelling the
``b'', ``c'', and ``d'' test planets orbiting a slow, medium, and
fast rotator. For comparison, Figure~\ref{fig:test_sys} further
shows the distributions obtained analysing the planets separately.
As one can expect, considering simultaneously multiple planets
narrows, and thus improves, the results. This can be best seen in
the middle panel (i.e., medium rotator), where the innermost
planet constrains the upper boundary of $P_{\rm rot}^{150}$, while
the outer planets constrain the lower boundary of $P_{\rm
rot}^{150}$, and in the right panel, for which the final
distribution is significantly narrower.

The mock system we used here has been constructed in such a way
that the uncertainties on the planetary parameters are homogeneous
among the three planets. However, this is unlikely to be the case
for real systems. Therefore, we expect that a meaningful
simultaneous analysis of multi-planet systems requires that the
masses and radii of the planets considered in the analysis have
sufficiently low uncertainties.

{The results obtained on multi-planet systems are not different
from those of single-planet systems. This is not surprising given
that those results correspond to the superposition of what is
separately obtained for each single planet in the system. This
indicates that for multi-planet systems, having one planet with a
good mass estimate is enough to constrain the stellar rotation
history, even if modeling all planets in a system simultaneously.}

\subsubsection{Constraining poorly known planetary masses in multi-planet systems}\label{sec:test:masses}

{Above, we have shown that for systems with well estimated
planetary masses a simultaneous analysis of all planets may better
constrain the stellar rotation history. Here we consider the case
in which one or more planets in a system have a poorly constrained
or unknown mass. To consider this case, we slightly modified the
MCMC algorithm as follows. We turned the poorly constrained or
unknown planetary mass into the retrieval parameter and set the
history of the stellar rotation period (more precisely $x$) as
prior, where its distribution has been derived employing the
planet with the well constrained mass.}

\begin{figure*}
  \includegraphics[width=\hsize]{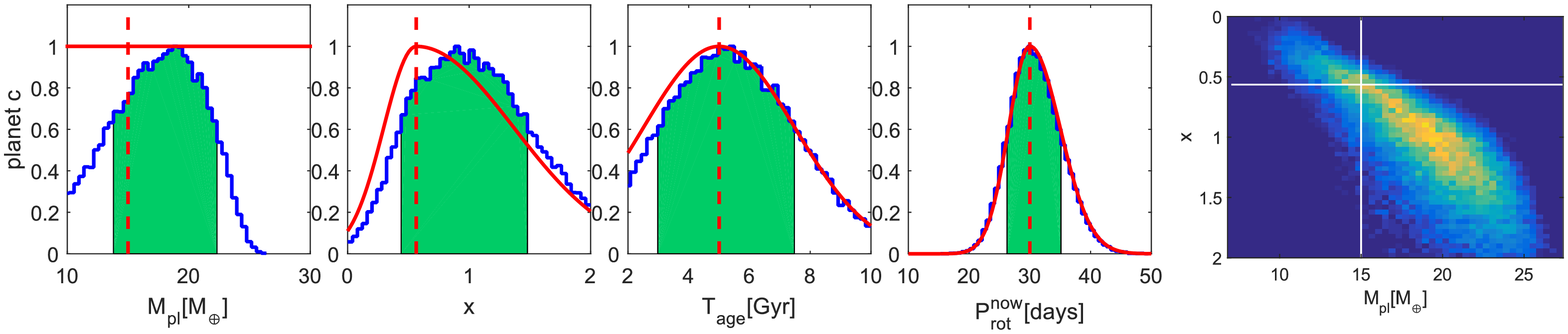}\\
  \includegraphics[width=\hsize]{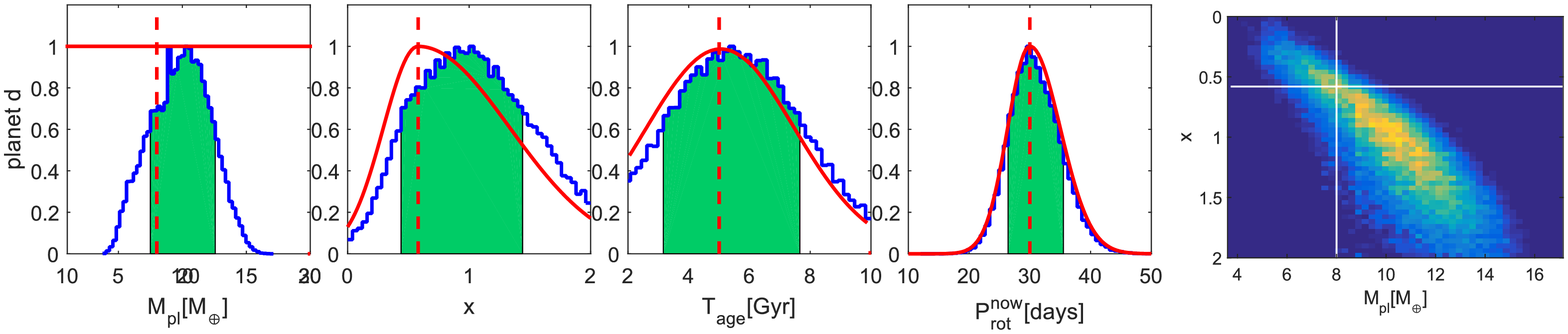}\\
  \caption{{Four leftmost columns: posterior distributions for planetary mass, $x$, age of the system, and present time stellar rotation period. Right column: pair-wise distribution for planetary mass and $x$. The red dashed lines in the four left panels and the white lines in the right panel show the injected values, while the solid red lines show the priors.
  The top line shows the results for the mock planet ``c'', and bottom line shows the results for planet ``d''.}}\label{fig:test_mass}
\end{figure*}

{Figure~\ref{fig:test_mass} shows as an example the results
obtained to recover the mass of the test planet ``c'' lying within
the mock system with a medium rotator, employing the value of the
exponent $x$ derived from the analysis of planet ``d'', and
otherwise, the results obtained for the mass of the planet ``d'',
assuming the exponent $x$ resolved for the planet ``c''. For both
test planets, we considered a uniform mass prior on the planetary
mass ranging between 1 and 39 Earth masses.
Figure~\ref{fig:test_mass} shows that the true masses lie within
the obtained 68$\%$ HPD credible intervals.}

\section{Application to Observed Planetary Systems}
\label{sec:results}
We now apply our atmospheric evolution framework to the HD3167 and
K2-32 planetary systems aiming at estimating the rotation rate
evolution of their host stars.  Table~\ref{tab:allplanets} lists
the adopted system parameters and results. {Although in
Section\,\ref{sec:test:multiplanet} we conclude that it is
possible to derive the stellar rotation history of the host star
with a simultaneous analysis of all planets in the system, here we
consider just one planet in each system. This is because for each
of the HD3167 and K2-32 planetary systems just one planet holds an
hydrogen atmosphere and/or has a mass which has been measured to
better than 30$\%$. For the K2-32 system, we use the $P_{rot}$
distribution obtained using the planet with a well-measured mass
to constrain the rotation history of the host star and the mass of
the other planets in the system (see
Section~\ref{sec:test:masses}).}

\subsection{HD~3167}
The HD3167 planetary system is composed of three sub-Neptune-mass
planets, where the innermost (HD3167b) and outermost (HD3167c)
planets transit their host star \citep{vanderburg2016,
gandolfi2017, christiansen2017}.  All planets in the system have
masses derived from radial-velocity measurements.
\begin{figure*}
\begin{centering}
\includegraphics[width=0.9\hsize]{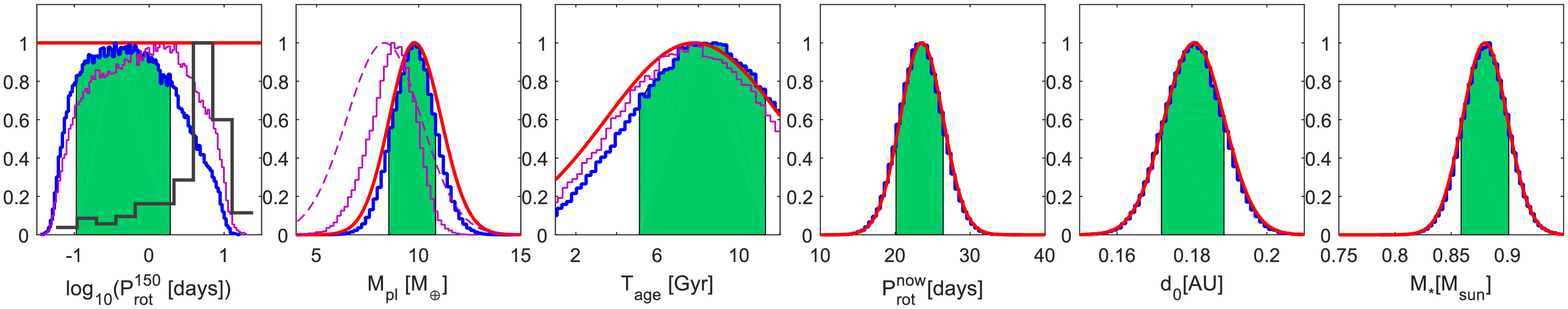}
\caption{MCMC posterior distributions {for $P_{\rm rot}^{150}$,
\Mpl, and system's age obtained from the modeling of HD3167c based
on the data from \citet{christiansen2017}.} The shaded areas
correspond to the 68\% HPD credible interval. The violet solid
lines show the same distributions obtained considering the system
parameters given by \citet{gandolfi2017}, while the violet dashed
line shows the assumed \Mpl\ prior. The black line histogram shows
the distribution of rotation periods for open cluster stars with
masses between 0.8 and 0.9\,$M_{\odot}$
\citep{johnstone2015rot}.} \label{fig:hd3167_c}
\end{centering}
\end{figure*}

HD3167b has a measured average density consistent with a rocky
composition, suggesting that the planet has lost its primordial
hydrogen-dominated atmosphere \citep{gandolfi2017}.  We indeed
find that the hydrogen envelope of HD3167b escapes completely
within the first 0.01\,Myr for any reasonable $x$ value (i.e.,
$0<x<2$). {We cannot use this planet for our analysis as it would
return a flat distribution of the stellar rotation period.}

HD3167c has an average density of $\sim$2\,g\,cm$^{-3}$,
suggesting that the planet still retains part of its primary
hydrogen-dominated atmosphere.  The evolutionary tracks do not
depend on the initial conditions, except for a slow rotator and a
planetary mass above 10.9\,\Me, for which the atmospheric
evolutionary tracks merge at $f_{\rm
at}$\,$\approx$\,7$\times$10$^{-2}$ and at an age of
$\approx$1\,Gyr. Figure~\ref{fig:hd3167_c} shows the posterior
distributions obtained adopting the system parameters given by
\citet{christiansen2017}, indicating that the host star was likely
to be a fast to moderate rotator. In particular, we obtain that at
150\,Myr the star emitted 38--128 times the XUV flux of the
current Sun, where the upper boundary corresponds to the
saturation limit. The pair-wise posterior distributions of the
parameters involved in the modeling are shown in Appendix
(Figure~\ref{fig:appx9}). We run the simulations again adopting
the system parameters of \citet{gandolfi2017} obtaining consistent
results, except for a slightly broader $P_{\rm rot}^{150}$
posterior distribution towards larger values (i.e., slower
rotator) due to the lower planetary mass. To demonstrate the
constraining power of our technique, Figure~\ref{fig:hd3167_c}
shows the results obtained from the analysis of HD3167c in
comparison with the distribution of stellar rotation periods at an
age of 150\,Myr measured for 0.8 and 0.9 solar mass stars members
of open clusters \citep{johnstone2015rot}.

HD3167d is not transiting and therefore does not have a measured
radius. Nevertheless, the stellar rotation rate evolutionary
tracks based on the results obtained from HD3167c lead us to
constrain the possible atmospheric evolution of HD3167d suggesting
that the planet should have completely lost its primary,
hydrogen-dominated atmosphere within 0.02--1.48\,Gyr.
\subsection{K2-32}
The K2-32 planetary system is composed of three transiting
sub-Neptune-mass planets.  Radial-velocity measurements placed
loose constraints on the planetary masses \citep{dai2016,
petigura2017}: K2-32c has an upper-mass limit, whereas for K2-32d
the mass is measured with a 50\% uncertainty. {The masses of these
planets are too poorly constrained to be useful for our analysis.}
Therefore, we focused our analysis on K2-32b, which has a low
average density of 0.67$\pm$0.16\,g\,cm$^{-3}$, strongly
indicative of the presence of an extended hydrogen envelope.  For
the analysis, we consider the system parameters given by
\citet{petigura2017}.
\begin{figure*}
\begin{centering}
\includegraphics[width=0.9\hsize]{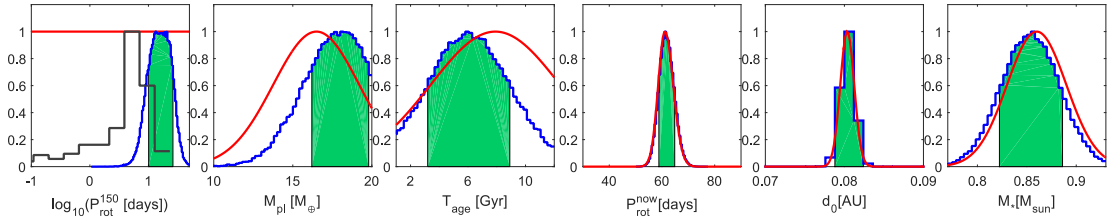}\\
\caption{Same as Figure~\ref{fig:hd3167_c}, but for K2-32b.}
\label{fig:k2032b}
\end{centering}
\end{figure*}

Our results (Figure~\ref{fig:k2032b}) strongly favor the slow
rotator scenario with the posterior distribution indicating
$P_{\rm rot}^{150}$ values lying between 10 and 26\,days, which
coincides with the slow rotation edge of the distribution given by
\citet{johnstone2015rot}, and corresponds to XUV fluxes at
150\,Myr between 0.5 and 4.0 times those of the present Sun.  This
means that, within our framework, the existence of such a
close-in, low-density planet is possible only if the star has
always been a very slow rotator. The pair-wise posterior
distributions of the parameters involved in the modeling are shown
in the Appendix (Figure~\ref{fig:appx10}).

Given the weak mass constraints for K2-32c and K2-32d, we used the
results on the stellar rotation rate history obtained from the
analysis of K2-32b to place further constraints on the estimated
planetary masses, {as described in Section~\ref{sec:test:masses}.}
The low measured average density of K2-32d
(1.38$^{+0.92}_{-0.67}$\,g\,cm$^{-3}$) suggests that the planet
has a significant hydrogen-dominated envelope. Assuming the range
of possible stellar rotation rates found from the analysis of
K2-32b, we expect the mass of K2-32d to range between 5.60 and
6.87\,\Me.

For K2-32c, by adopting a mass equal to the upper mass limit and
the results on the stellar rotation rate derived from K2-32b, we
obtain a planetary radius significantly larger than the measured
one, with the planet retaining nearly 50\% of its initial
envelope.  To fit the observed \Rpl, we require planetary masses
ranging between 4.55 and 7.29\,\Me.  By adopting instead the
planetary radius given by \citet{mayo2018}, the range of possible
planetary masses shifts slightly towards larger values, namely
4.76--7.52\,\Me.

{We further test our results by carrying out an additional MCMC
run including all three planets. With this, we aim at confirming
that simultaneously considering all three planets does not affect
the results on the stellar rotation history, and test if the
posterior distributions for the masses of K2-32c and K2-32d are
consistent with the above estimates. The results of this test are
shown in Figure~\ref{fig:appx11}. As expected, the $P_{\rm
rot}^{150}$ distribution is identical to that obtained considering
only K2-32b. Furthermore, the 68$\%$ HPD credible intervals for
the masses of planets c and d lie between 6.16 and 7.68\,\Me\ and
between 5.36 and 6.73\,\Me, respectively. Except for a slight
difference in the lower mass boundary for planet c, these values
are comparable to those obtained using the algorithm described in
Section~\ref{sec:test:masses}.}

%
\section{Discussions and conclusions}
\label{sec:conclusion}
The currently observed atmospheric properties of exoplanets
strongly depend on their mass-loss history, which can reveal the
otherwise unknown evolution history of the stellar high-energy
radiation.  We developed a framework to compute planetary radii as
a function of time and coupled it to a Bayesian sampler to infer
the rotation rate history of the host star, which is constrained
by the measured basic system parameters.  We applied this
framework to the HD3167 and K2-32 systems.

Despite having very similar masses, we find that the K-type stars
HD3167 and K2-32 had a widely different initial rotation-rate,
thus high-energy emission.  Most likely, HD3167 evolved as a
medium (or fast) rotator, while K2-32 evolved as an extremely slow
rotator.  We further showed that our evolution model can lay
further constraints for planetary systems in which some of the
planetary parameters are poorly measured. {For the K2-32 system,
the stellar rotation rate constrained by K2-32b, leads to masses
ranging between 6.16 and 7.68\,{\Me} for K2-32c, and masses
ranging between 5.36 and 6.73\,{\Me} for K2-32d.}

{We found that in comparison to single-planet systems,
multi-planet systems have a major advantage. Multi-planet systems
containing more than one planet with well measured mass enable to
more precisely constraint the stellar rotation history.
Multi-planet systems containing one planet with well-measured mass
and other planets with poorly measured masses our technique
enables one to use the former to derive the stellar rotation
history, which can then be used to constrain the mass of the
latter planets.}

We have shown that atmospheric evolution modeling of close-in
planets has the power to uncover the past rotation history of
late-type stars.  It enables one also to constrain poorly known
parameters (i.e., mass and/or radius) of some of the planets in a
system. These results can be extended to other systems provided
that they host a close-in planet with a mass close to that of
Neptune or smaller and with a hydrogen-dominated envelope. A
systematic analysis of a large sample of systems will also help to
set constraints on the different model assumptions adopted by this
framework.  Furthermore, by comparing the output stellar rotation
period distribution at a given age with what is measured in
clusters, or by comparing the results obtained from two planets in
a system, it would be possible to identify planets that evolved
differently from what expected. These planets might have been
affected, for example, by strong impacts, or experienced an early
dispersal of the protoplanetary nebula, or underwent strong
hydrogen release from the solidifying rocky core
\citep[e.g.,][]{chachan2018,bonomo2019}. Firmly identifying these
``odd'' systems would be the first step towards studying them to
better understand the evolution of planetary systems.

Our results show also the importance of precisely constraining
system parameters, particularly the system's age and planetary
mass and radius. The ESA {\it PLATO} mission \citep{rauer2014}
will be in the best position to simultaneously discover transiting
planetary systems and provide system parameters with the necessary
accuracy to pin down the evolutionary track of the planetary
atmosphere and stellar rotation rate.
\begin{acknowledgements}
We acknowledge the FFG project P853993, the FWF/NFN projects
S11607-N16 and S11604-N16, and the FWF projects P27256-N27 and
P30949-N36. N.V.E. acknowledges support by the Russian Science
Foundation grant No 18-12-00080. We thank the anonymous referee
for insightful comments.
%
\end{acknowledgements}
%

%

\begin{rotatetable*}
\begin{deluxetable*}{c|c|c|c|c|c|c|c|c|c}
\tablecaption{ \label{tab:allplanets} {Adopted system parameters
and data, and obtained results ($x$) for HD3167 and K2-32.}  The
cases for which \Rpl\ reaches the core radius are given in
parenthesis.}

\tabletypesize{\scriptsize}
\tablehead{
\colhead{\hspace{1 cm} planet} & \colhead{$M_*$}   & \colhead{age}  &
\colhead{$P_{\rm rot}$}         & \colhead{\Mpl}    & \colhead{\Rpl} &
\colhead{$\rho$}               & \colhead{$d_0$}   & \colhead{$\Lambda$} &
\colhead{$x$} \\
\colhead{}       & \colhead{[\Mo]} & \colhead{[Gyr]} &
\colhead{[days]} & \colhead{[\Me]} & \colhead{[\Re]} & \colhead{[g/cm$^3$]} &
\colhead{[AU]}   & \colhead{}      & \colhead{}}
\startdata
HD3167b & $0.88\pm0.02$ & $7.8\pm4.3$ & $23.52\pm2.87$ & $5.69\pm0.44$ (a) & $(1.57\pm0.05)$ (a)       & $8.00^{+1.10}_{-0.98}$ & $0.01752\pm0.00063$ & $15.1$ & -- \\
            &               &             &                & $5.02\pm0.38$ (b) & $(1.70^{+0.18}_{-0.15})$ (b) & $5.60^{+2.15}_{-1.43}$ &                     & $12.3$ &    \\
            &               &             &                &                   & $(1.56\pm0.06)$ (c)       &   &   &   &  \\
            &               &             &                &                   & $(1.62^{+0.17}_{-0.08})$ (d) &   &   &   &  \\
HD3167c & $0.88\pm0.02$ & $7.8\pm4.3$ & $23.52\pm2.87$ & $8.33^{+1.79}_{-1.85}$ (a) & $2.74^{+0.11}_{-10}$ (a) & $2.21^{+0.56}_{-0.53}$ & $0.1806\pm0.0080$ &  $39.7$  & $0.63^{+1.14}_{-0.08}$ \\
            &   &   &   & $9.80^{+1.30}_{-1.24}$ (b) & $3.01^{+0.42}_{-0.28}$ (b) & $1.97^{+0.94}_{-0.59}$ &   & $42.6$  &  $1.06^{+0.86}_{-0.26}$  \\
            &   &   &   &   & $2.85^{+0.24}_{-0.15}$ (c) &   &   &   &    \\
            &   &   &   &   & $2.83^{+0.31}_{-0.14}$ (d) &   &   &   &    \\
HD3167d & $0.88\pm0.02$ & $7.8\pm4.3$   & $23.52\pm2.87$       & $6.90\pm0.71$ (b)  & $(1.90 \pm 0.07)^*$ & --            & $0.07757\pm0.00027$ & -- & -- \\
K2-32b    & $0.86\pm0.03$ & $7.90\pm4.50$ & $61.10^{+3.18}_{-2.53}$ & $16.50\pm2.70$ (e) & $5.13\pm0.28$ (e)   & $0.67\pm0.16$ & $0.08036\pm0.00088$ & $29.1$ & $0.09_{-0.07}^{+0.34}$ \\
         &   &   &   & $21.10\pm5.90$ (f) & $5.13\pm0.28$ (f)       & $0.67\pm0.16$ &   & $37.2$  &     \\
         &   &   &   &                    & $5.62\pm0.45$ (g)       &   &   &   &    \\
         &   &   &   &                    & $5.17^{+0.20}_{-0.16}$ (d) &   &   &   &    \\
K2-32c & $0.86\pm0.03$ & $7.90\pm4.50$ & $61.10^{+3.18}_{-2.53}$ & $<12.10$ (e) & $3.01\pm0.25$ (e) & $<2.70$ & $0.1399\pm0.0015$ &  $<48.0$ & -- \\
         &   &   &   & $<8.1$ (f) & $3.48^{+0.97}_{-0.42}$ (f) & $<1.10$ &      &  $<27.8$  &   \\
         &   &   &   &   & $3.32\pm0.33$ (g) &   &   &   &     \\
         &   &   &   &   & $3.12^{+0.18}_{-0.12}$ (d) &   &   &       &    \\
K2-32d & $0.86\pm0.03$ & $7.90\pm4.50$ & $61.10^{+3.18}_{-2.53}$ & $10.30\pm4.70$ (e) & $3.43\pm0.35$ (e) & $1.38^{+0.92}_{-0.67}$ & $0.1862\pm0.0020$ & $41.6$ & -- \\
         &   &   &   & $<35.00$ (f) & $3.75\pm0.40$ (f) & $<3.60$ &   & $<129$ &     \\
         &   &   &   &   & $3.77\pm0.41$ (g) &   &   &   &    \\
         &   &   &   &   & $3.41^{+0.26}_{-0.14}$ (d) &   &   &   &    \\
\enddata
\tablenotetext{}{
a -- \citet{gandolfi2017}, b -- \citet{christiansen2017},
c -- \citet{vanderburg2016}, d -- \citet{mayo2018},
e -- \citet{petigura2017}, f -- \citet{dai2016},
g -- \citet{crossfield2016}, h -- \citet{weiss2013},
i -- \citet{morton2016}, j -- \citet{berger2018},
k -- \citet{masuda2013}, l -- \citet{hadden2017}. \\
$^*$ -- the {radius derived in} this paper assuming the rotation
rate obtained for HD~3167\,c.}
\end{deluxetable*}
\end{rotatetable*}

\begin{appendix}

\begin{figure}[hb!]
\begin{center}
  \includegraphics[width=0.8\hsize]{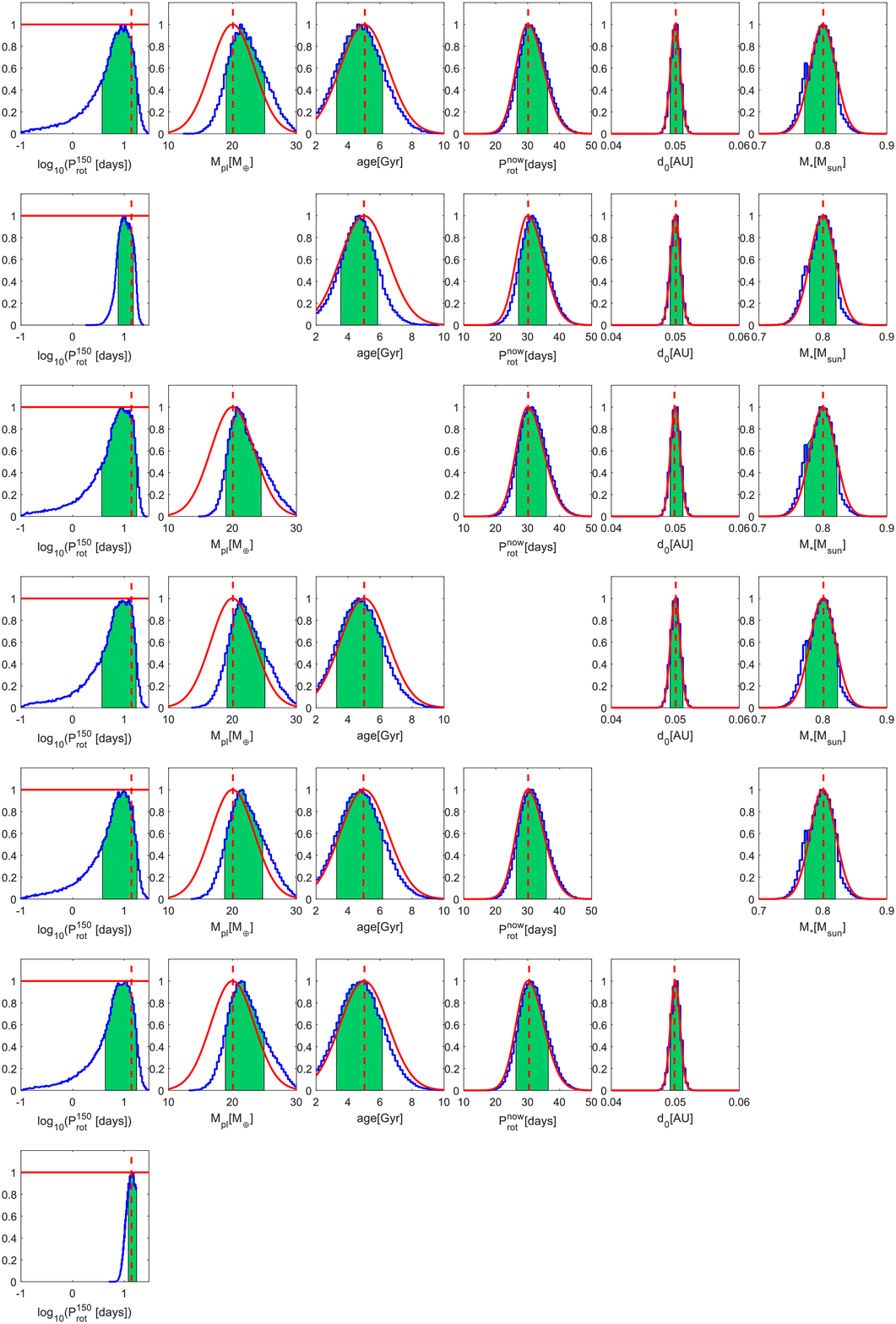}\\
  \caption{Same as Figure~\ref{fig:test_hist}, but for the test planet ``b'' orbiting a slow rotator.}\label{fig:appx1}
\end{center}
\end{figure}

\begin{figure}
\begin{center}
  \includegraphics[width=0.8\hsize]{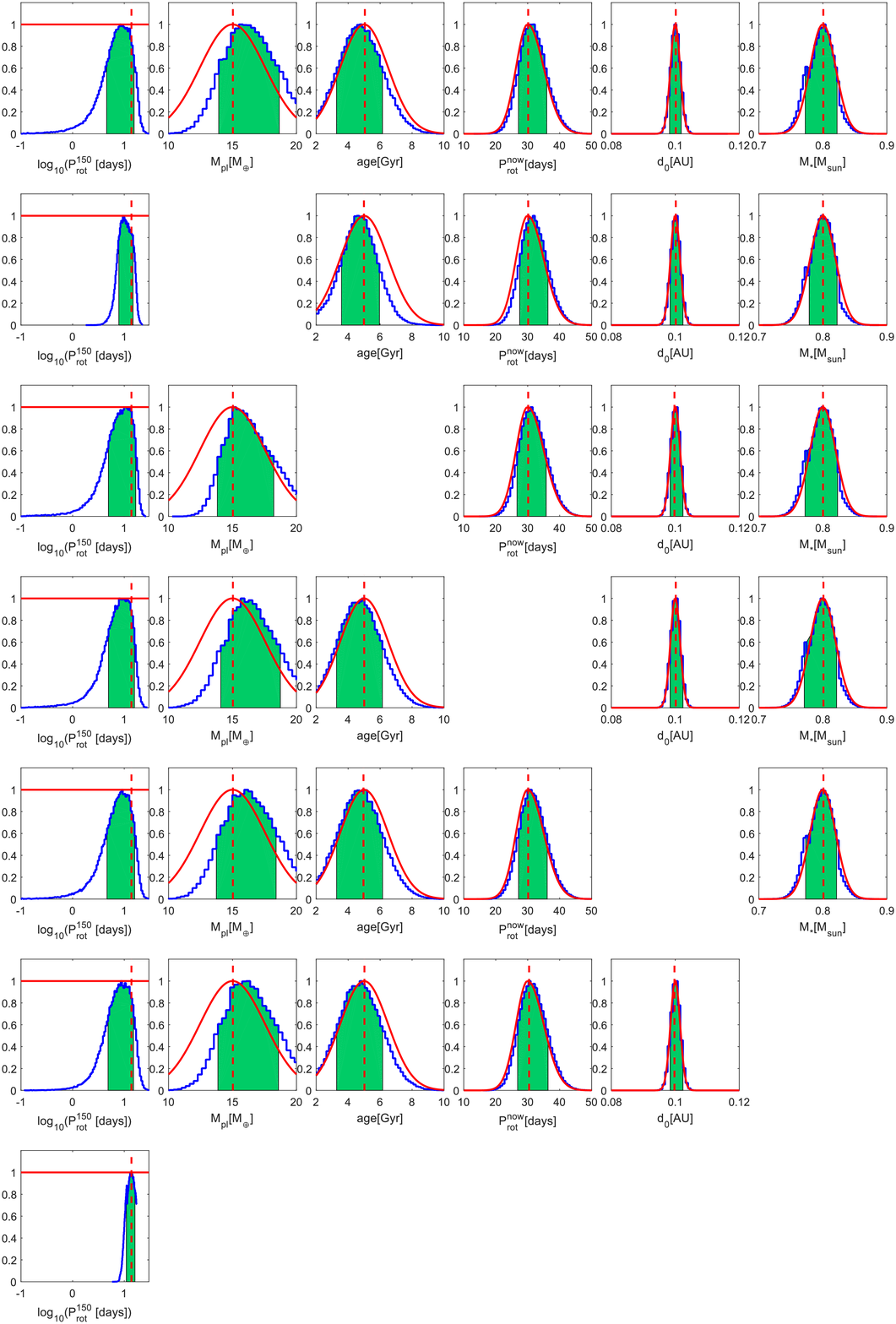}\\
  \caption{Same as Figure~\ref{fig:test_hist}, but for the test planet ``c'' orbiting a slow rotator.}\label{fig:appx2}
\end{center}
\end{figure}

\begin{figure}
\begin{center}
  \includegraphics[width=0.8\hsize]{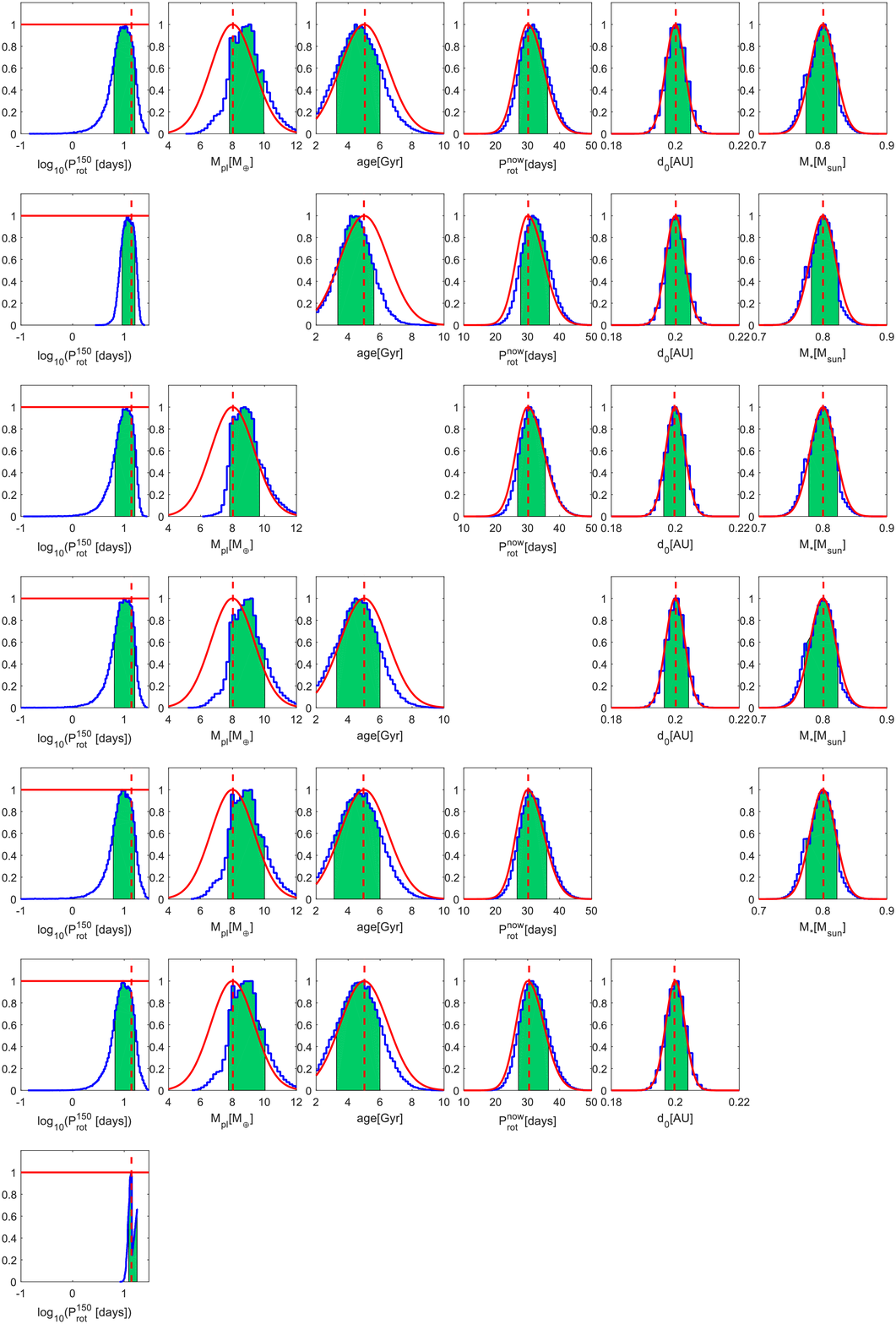}\\
  \caption{Same as Figure~\ref{fig:test_hist}, but for the test planet ``d'' orbiting a slow rotator.}\label{fig:appx3}
\end{center}
\end{figure}

\begin{figure}
\begin{center}
  \includegraphics[width=0.8\hsize]{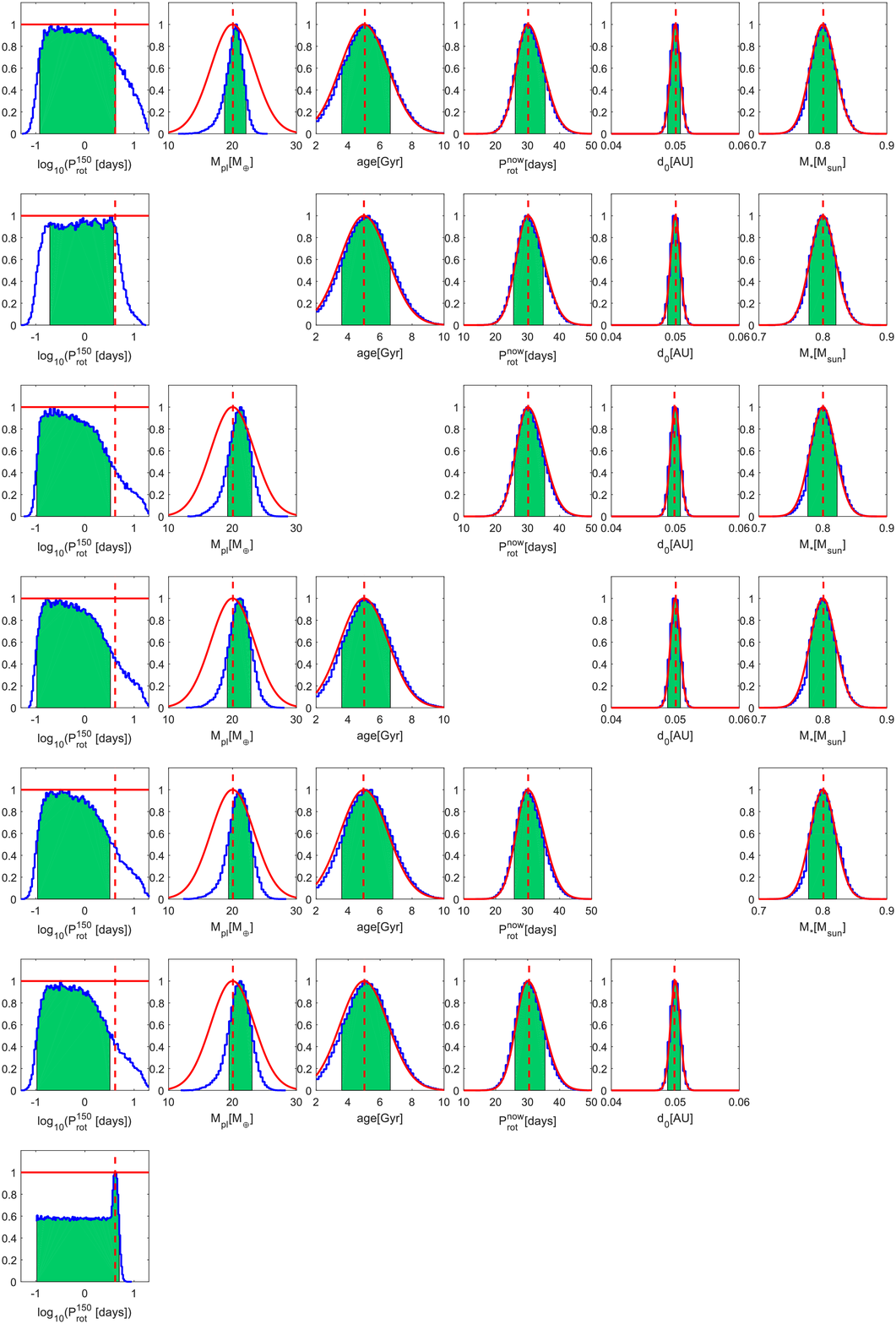}\\
  \caption{Same as Figure~\ref{fig:test_hist}, but for the test planet ``b'' orbiting a medium rotator.}\label{fig:appx4}
\end{center}
\end{figure}

\begin{figure}
\begin{center}
  \includegraphics[width=0.8\hsize]{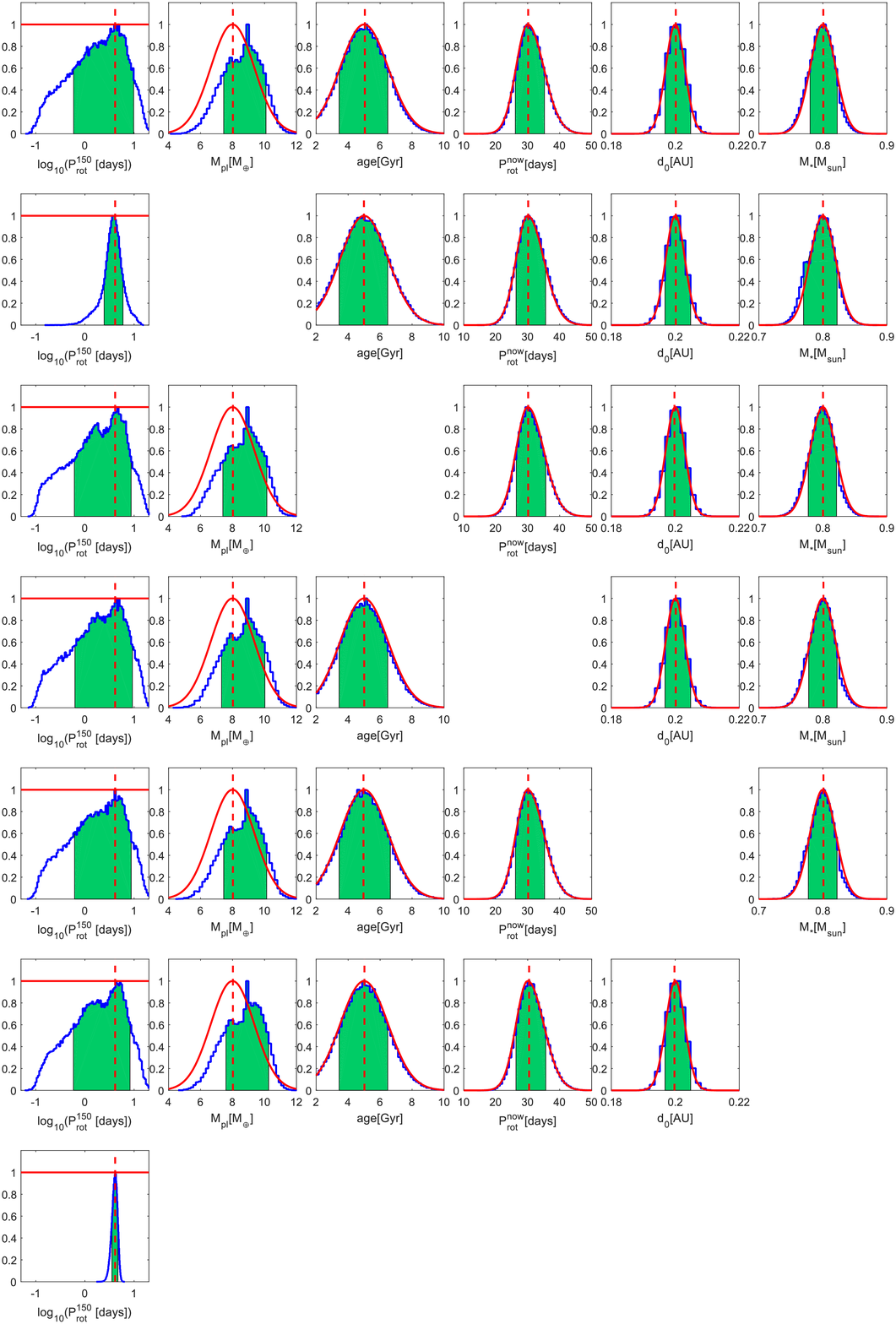}\\
  \caption{Same as Figure~\ref{fig:test_hist}, but for the test planet ``d'' orbiting a medium rotator.}\label{fig:appx5}
\end{center}
\end{figure}

\begin{figure}
\begin{center}
  \includegraphics[width=0.8\hsize]{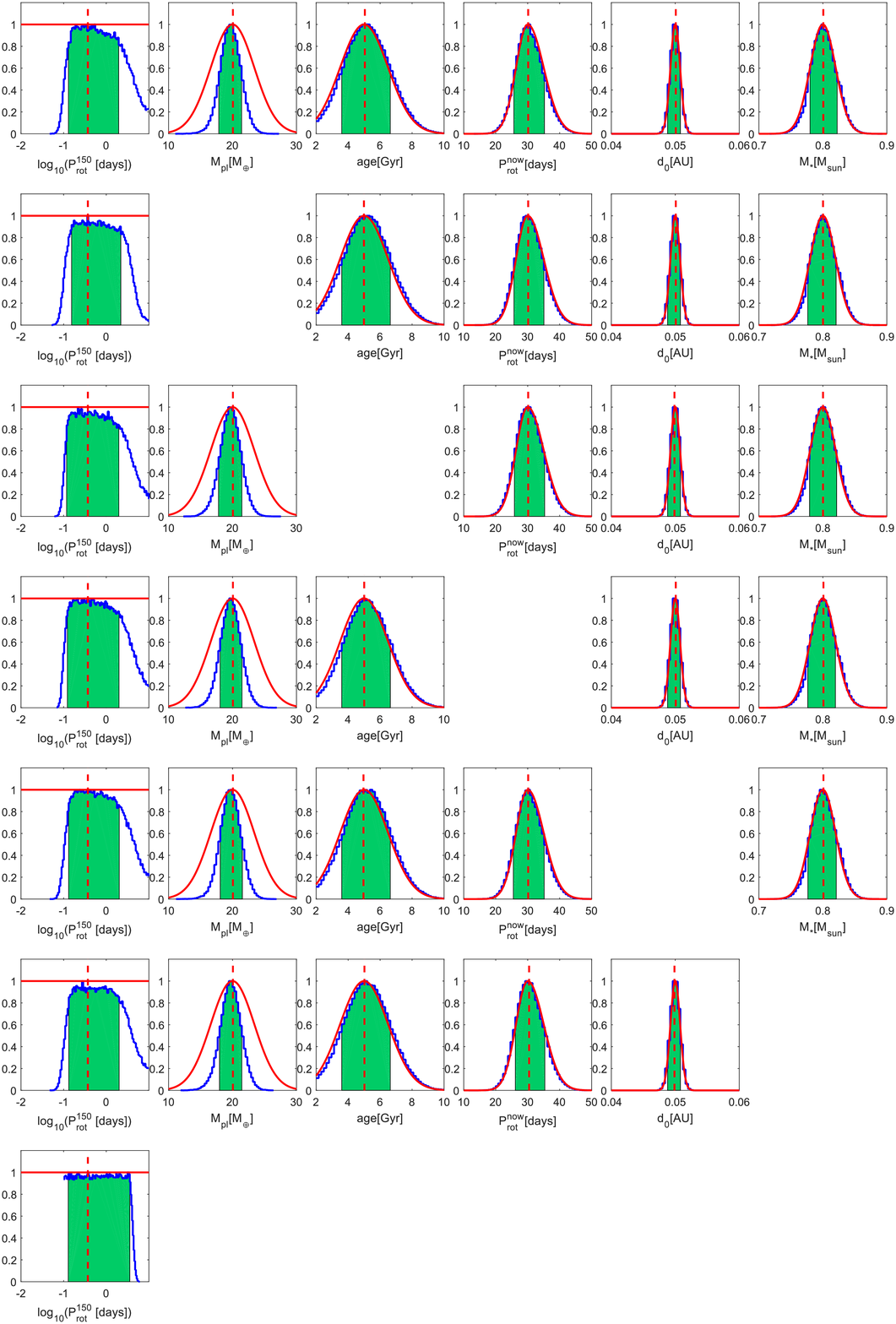}\\
  \caption{Same as Figure~\ref{fig:test_hist}, but for the test planet ``b'' orbiting a fast rotator.}\label{fig:appx6}
\end{center}
\end{figure}

\begin{figure}
\begin{center}
  \includegraphics[width=0.8\hsize]{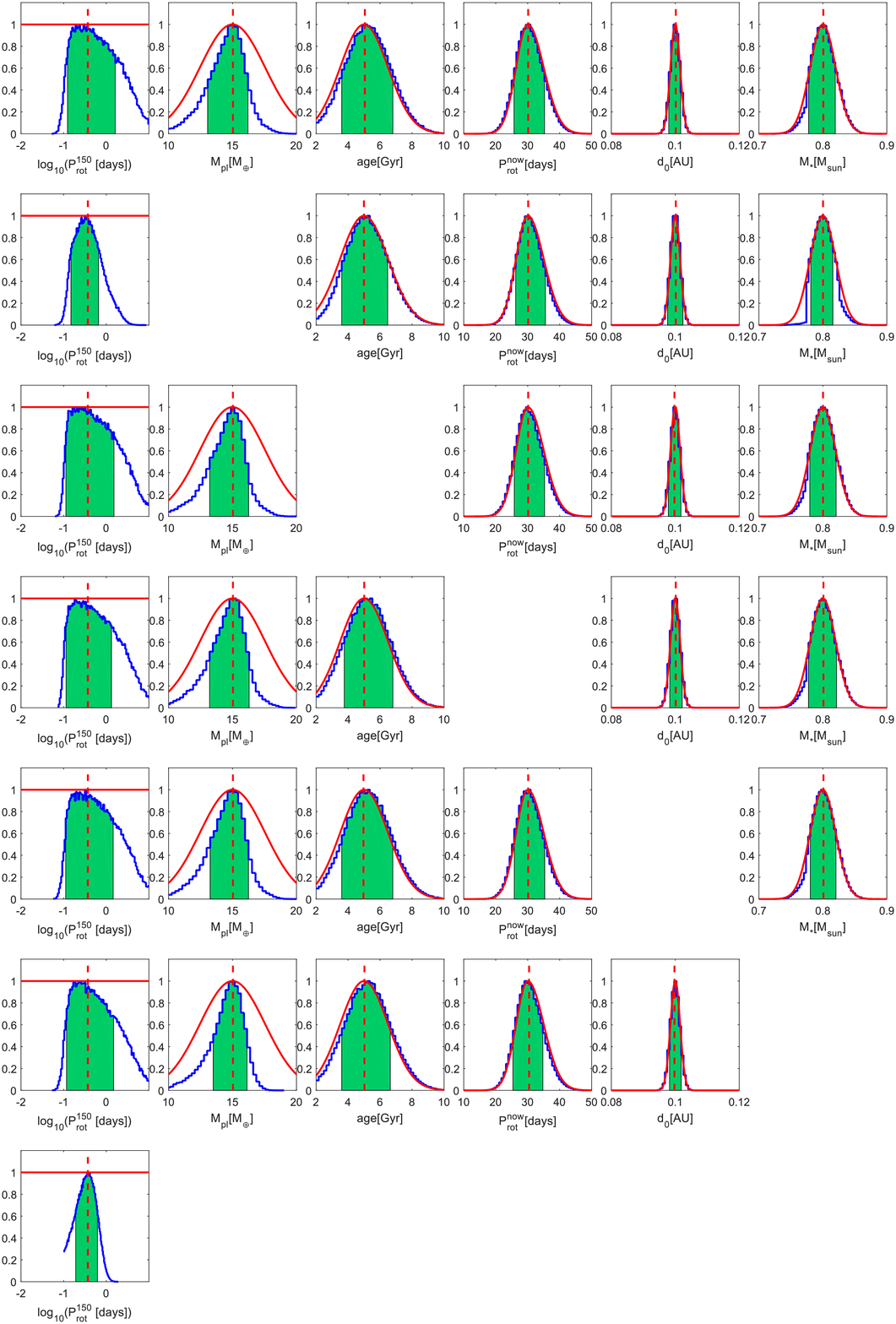}\\
  \caption{Same as Figure~\ref{fig:test_hist}, but for the test planet ``c'' orbiting a fast rotator.}\label{fig:appx7}
\end{center}
\end{figure}

\begin{figure}
\begin{center}
  \includegraphics[width=0.8\hsize]{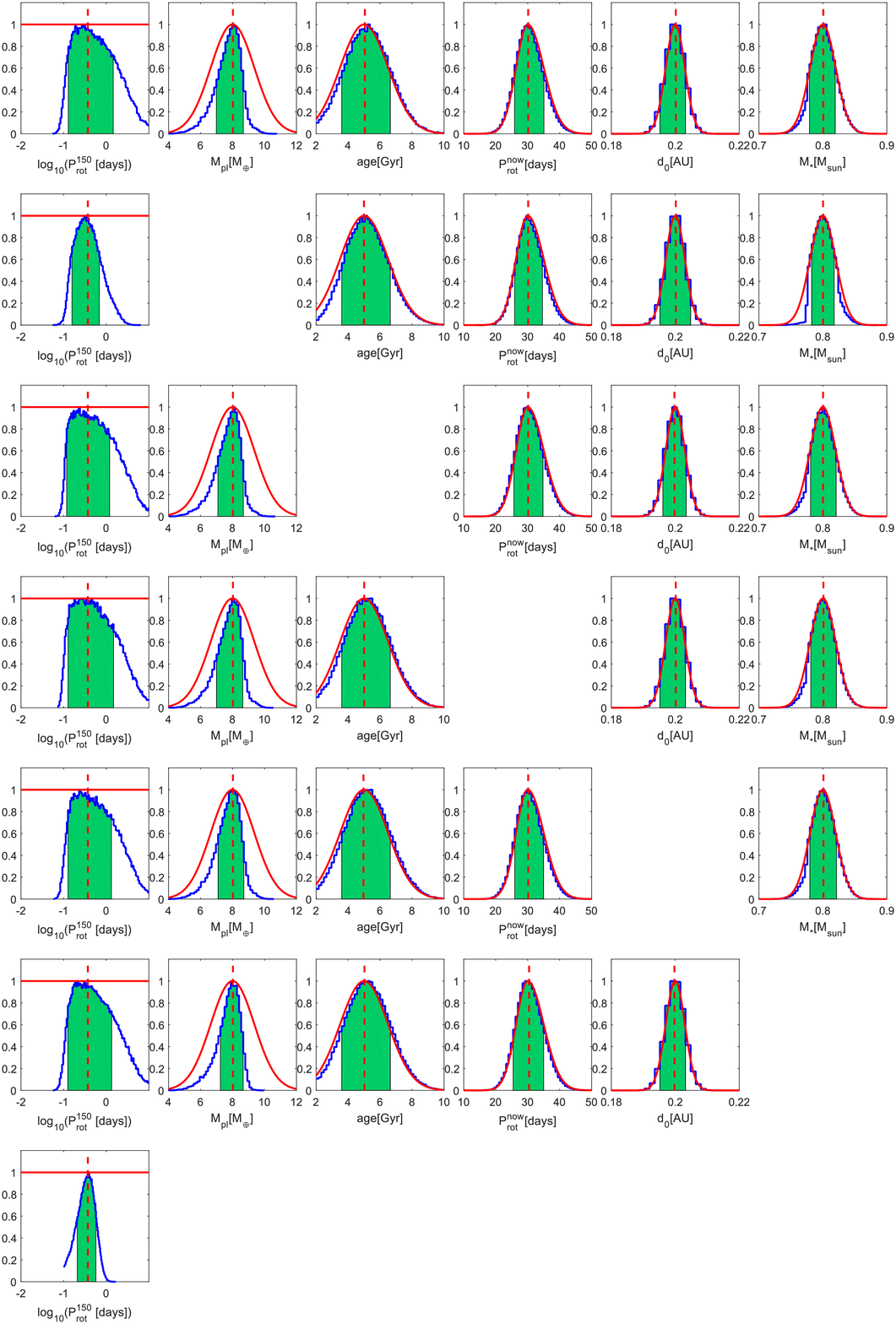}\\
  \caption{Same as Figure~\ref{fig:test_hist}, but for the test planet ``d'' orbiting a fast rotator.}\label{fig:appx8}
\end{center}
\end{figure}

\begin{figure}
\begin{center}
  \includegraphics[width=0.8\hsize]{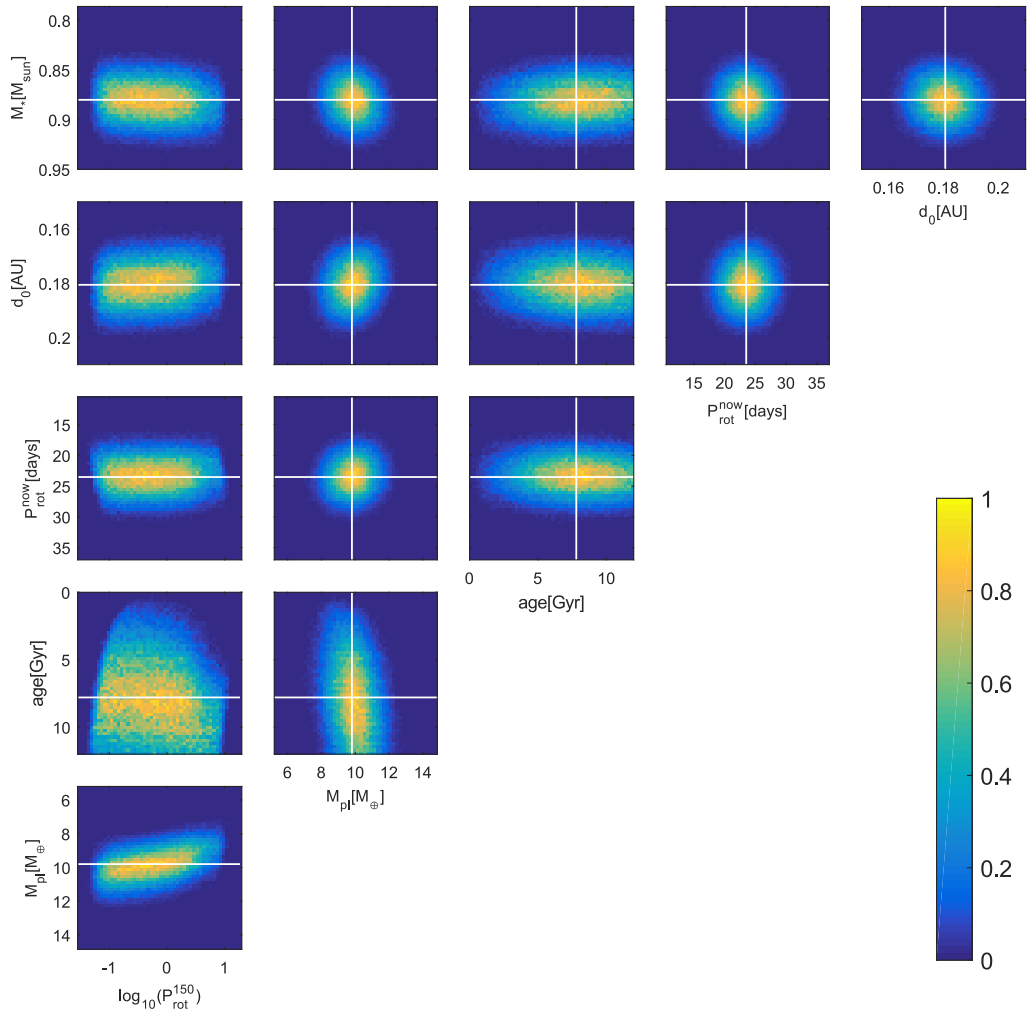}\\
  \caption{Pair-wise distributions for the parameters involved in the modeling of HD3167c.}\label{fig:appx9}
\end{center}
\end{figure}

\begin{figure}
\begin{center}
  \includegraphics[width=0.8\hsize]{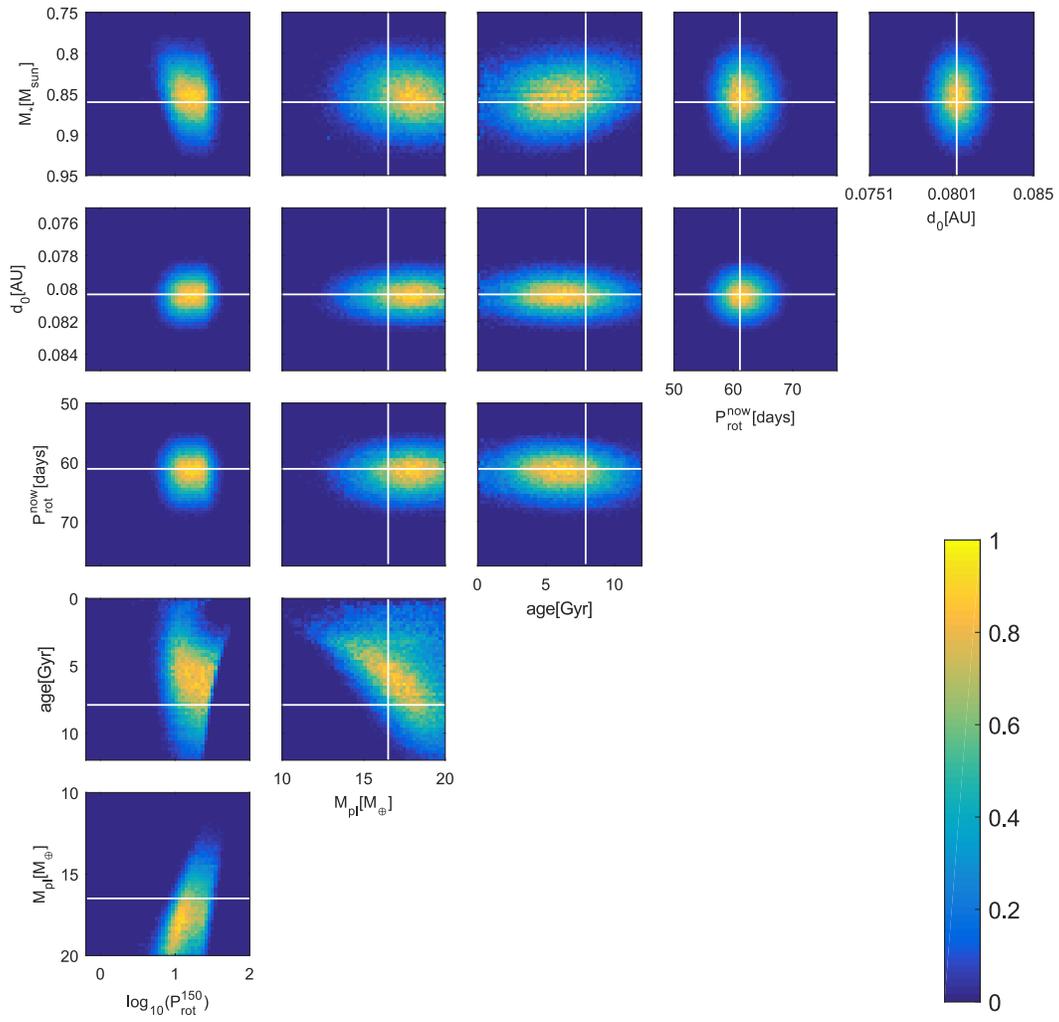}\\
  \caption{Pair-wise distributions for the parameters involved in the modeling of K2-32b.}\label{fig:appx10}
\end{center}
\end{figure}

\begin{figure}
\begin{center}
  \includegraphics[width=0.8\hsize]{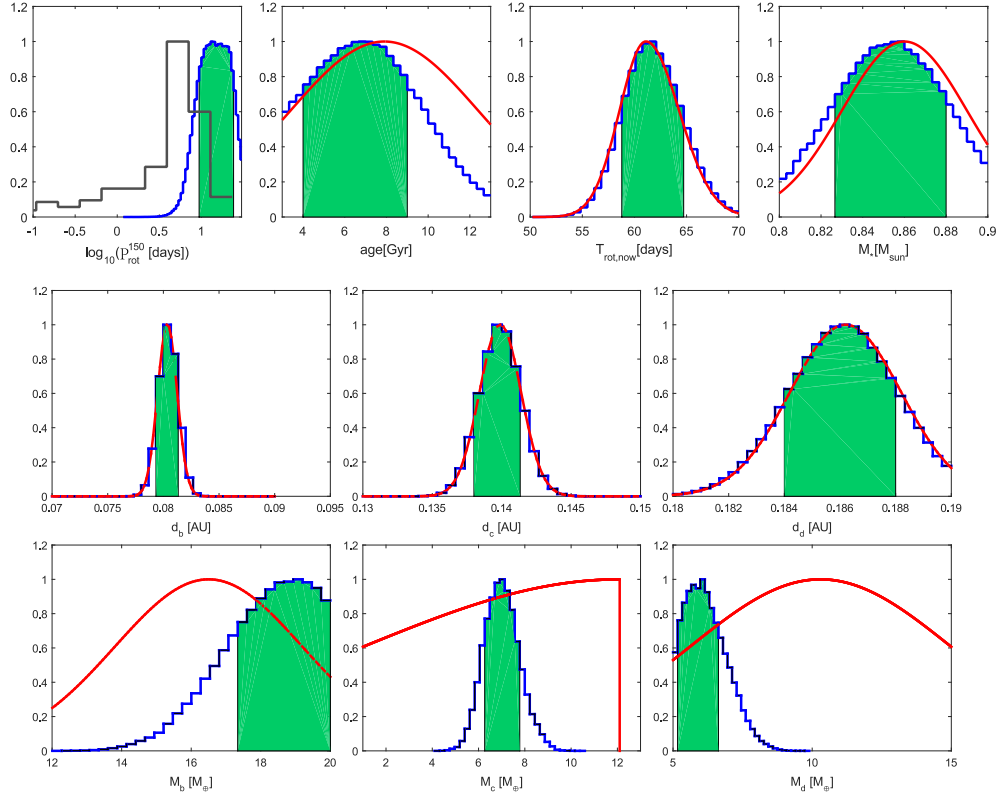}\\
  \caption{MCMC posterior distributions for $P_{\rm rot}^{150}$,
system's age, present time stellar rotation period, stellar mass,
planetary orbital separations, and masses obtained from
the joint modeling for the K2-32 system. The shaded areas
correspond to the 68\% HPD credible interval. The black line
histogram shows the distribution of rotation periods obtained from open
cluster stars with masses between 0.8 and 0.9\,$M_{\odot}$
\citep{johnstone2015rot}.}\label{fig:appx11}
\end{center}
\end{figure}

\end{appendix}

\end{document}